\documentclass[12pt,preprint]{aastex}
\usepackage{xspace}
\usepackage{graphicx}
\usepackage{natbib}
\newcommand{\nraoblurb}{The National Radio Astronomy Observatory is
a facility of the National Science Foundation operated under cooperative
agreement by Associated Universities, Inc.}
\newcommand{\hide}[1]{}

\newcommand{\expo}[1]{\ensuremath{10^{#1}}\xspace}

\newcommand{\gsim}{\ensuremath{\,\gtrsim\,}\xspace}
\newcommand{\lsim}{\ensuremath{\,\lesssim\,}\xspace}
\newcommand{\gl}{\ensuremath{\ell}\xspace}
\newcommand{\gb}{\ensuremath{{\it b}}\xspace}
\newcommand{\absb}{\ensuremath{\vert\,\gb\,\vert}\xspace}

\newcommand{\lb}{\ensuremath{(\gl,\gb)}\xspace}
\newcommand{\lv}{\ensuremath{(\gl,v)}\xspace}

\newcommand{\ra}{\ensuremath{\alpha}\xspace}
\newcommand{\dec}{\ensuremath{\delta}\xspace}
\newcommand{\radec}{\ensuremath{(\ra,\dec)}\xspace}

\newcommand{\kms}{\ensuremath{\,{\rm km\,s^{-1}}}\xspace}

\newcommand{\m}{\ensuremath{\,{\rm m}}\xspace}
\newcommand{\cm}{\ensuremath{\,{\rm cm}}\xspace}

\newcommand{\kpc}{\ensuremath{\,{\rm kpc}}\xspace}
\newcommand{\K}{\ensuremath{\,{\rm K}}\xspace}
\newcommand{\mK}{\ensuremath{\,{\rm mK}}\xspace}

\newcommand{\mhz}{\ensuremath{\,{\rm MHz}}\xspace}
\newcommand{\ghz}{\ensuremath{\,{\rm GHz}}\xspace}

\newcommand{\degree}{\ensuremath{\,^\circ}\xspace}

\newcommand{\mjy}{\ensuremath{\,{\rm mJy}}\xspace}

\newcommand{\te}{\ensuremath{{T_{e}}}\xspace}

\newcommand{\rgal}{\ensuremath{\,R_G}\xspace}   
\newcommand{\dsun}{\ensuremath{\,d_\sun}\xspace}          

\newcommand{\hi}{{\rm H\,{\footnotesize I}}\xspace}
\newcommand{\hii}{{\rm H\,{\footnotesize II}}\xspace}

\newcommand{\hal}[1]{\ensuremath{{\rm H}\,{#1}\,\alpha\xspace}}
\newcommand{\hna}{\ensuremath{{\rm H}\,{\rm n}\,\alpha\xspace}}
\newcommand{\hnaa}{\ensuremath{\langle\,\hna\,\rangle\xspace}}

\def\tmbcom#1{{\bf TMB: #1}}

\shorttitle{GBT \hii\ Region Survey}
\shortauthors{Anderson et al.}

\begin{document}

\title{The Green Bank Telescope H\,{\small \bf II} Region Discovery Survey\\
II. The Source Catalog}

\author{L.~D.~Anderson\altaffilmark{1,2}, T.~M.~Bania\altaffilmark{1}, 
Dana~S.~Balser\altaffilmark{3}, \& Robert~T.~Rood\altaffilmark{4}}

\altaffiltext{1}{Institute for Astrophysical Research, Department of Astronomy,
Boston University, 725 Commonwealth Ave., Boston MA 02215, USA.}
\altaffiltext{2}{Current address: Laboratoire d'Astrophysique de Marseille (UMR
6110 CNRS \& Universit\'e de Provence), 38 rue F. Joliot-Curie, 13388 Marseille
Cedex 13, France.}
\altaffiltext{3}{National Radio Astronomy Observatory, 520 Edgemont Road,
Charlottesville VA, 22903-2475, USA.}
\altaffiltext{4}{Astronomy Department, University of Virginia, P.O. Box 3818,
Charlottesville VA 22903-0818, USA.}

\begin{abstract}

The Green Bank Telescope \hii\ Region Discovery Survey 
has doubled the number of known \hii\ regions in the
Galactic zone $343\,\arcdeg \le \gl \le 67\,\arcdeg$ with 
$\absb \le 1\,\arcdeg$.  
We detected 603 discrete hydrogen radio recombination line (RRL)
components at 9\ghz (3\cm) from 448 targets.  Our targets were
selected based on spatially coincident mid-infrared and 20\,cm radio
continuum emission.  Such sources are almost invariably \hii\ regions;
we detected hydrogen RRL emission from 95\% of our target sample.  The
sensitivity of the Green Bank Telescope and the power of its
spectrometer together made this survey possible.  Here we provide a
catalog of the measured properties of the RRL and continuum emission
from the survey nebulae.  The derived survey completeness limit,
180\,mJy at 9\ghz, is sufficient to detect all \hii\ regions ionized
by single O-stars to a distance of 12\,kpc.
These recently discovered nebulae share the same distribution on the
sky as does the previously known census of Galactic \hii\ regions.  On
average, however, the new nebulae have fainter continuum fluxes,
smaller continuum angular sizes, fainter RRL intensities and smaller
RRL line widths.  Though small in angular size, many of our new
nebulae show little spatial correlation with tracers associated with
extremely young \hii\ regions, implying that our sample spans a range
of evolutionary states.
We discovered 34 first quadrant negative-velocity \hii\ regions, which
lie at extreme distances from the Sun and appear to be part of the
Outer Arm.  We found RRL emission from 207 {\it Spitzer} GLIMPSE
8.0\,\micron\ ``bubble'' sources, 65 of which have been cataloged
previously.  It thus appears that nearly all GLIMPSE bubbles are
\hii\ regions and that $\sim50\%$ of all Galactic \hii\ regions have a
bubble morphology at 8.0\,\micron.

\end{abstract}



\section{Introduction\label{sec:intro}}

\hii\ regions are zones of plasma surrounding massive O- and B-type
stars that are kept ionized by the stellar far ultraviolet (FUV)
radiation.  Stars capable of creating \hii\ regions have lifetimes of
$\sim\,10$ million years.  These nebulae are zero age objects compared
to the age of the Milky Way and are therefore located at sites of
recent massive star formation. \hii\ regions are the most luminous
objects in the Milky Way at mid-infrared (MIR) to radio wavelengths
and can be seen across the entire Galactic disk.  Because of their
luminosity, they are the clearest indicators of the active sites of
massive star formation.  They are the archetypical tracers of Galactic
spiral structure. Their chemical abundances, moreover, indicate the
present state of the interstellar medium (ISM) and reveal the
elemental enrichment caused by the nuclear processing of many stellar
generations.  They provide unique and important probes of billions of
years of Galactic chemical evolution (GCE).

Knowing the Galactic distribution, kinematics, and physical properties
of star cluster/\hii region/PDR/molecular cloud complexes informs a
host of research areas, including the merger history of the Milky Way,
the structure of the Galaxy, the physics of star formation, and the
evolution of the ISM.  At the most basic level, a census locates new
sites of massive star formation and paves the way for future studies
at different wavelengths.  Individual Galactic \hii\ regions are
astrophysically important objects that reveal details of the impact of
the star formation process on the ISM.

\hii\ region RRL spectra yield the velocity of the nebulae, which can
then be used to derive their kinematic distances.  The distances place
\hii\ regions onto a Galactic map that traces Galactic structure
\citep[e.g.][]{downes80, anderson09a}.  The RRL line width measures a
combination of the turbulent and thermal energies in the \hii\ region
plasma and is a parameter needed to derive the electron temperature of
\hii\ regions.  The electron temperature is a proxy for nebular
metallicity and hence is key to understanding the Milky Way
metallicity gradient, which provides an important constraint on models
for GCE \citep[][]{wink83, shaver83, quireza06b}.

Radio continuum observations of \hii\ regions measure the free-free
thermal emission and give the radio flux and angular size.  When
combined with a distance estimate, these parameters give a luminosity
and a physical size. The former constrains the spectral type of the
ionizing star and the latter is related to the evolutionary stage of
the \hii\ region.  Finally, with a large volume-limited sample, we may
better estimate the number of \hii\ regions in the Galaxy and derive
their luminosity function.  Studies of external galaxies have found
that the shape of the \hii\ region luminosity distribution is related
to the morphology of the host Galaxy \citep[see][and references
  therein]{oey98}.  By deriving the slope of the luminosity
distribution of Galactic \hii\ regions, we may get a better
understanding of the morphology of the Milky Way.

Modern Galactic \hii\ region surveys began with studies of the Palomar
Sky Survey optical plates \citep{sharpless53, sharpless59}.
\citet{gum55} found 85 nebulae visible from the Southern hemisphere.
The Sharpless and Gum surveys were soon augmented by \citet{rcw}, who
cataloged 182 optical \hii\ regions in the Southern sky.  (Some of
these are also Sharpless nebulae.)  Extinction within the Galactic
plane, however, limits the detection of optical nebulae to within a
few kpc of the Sun.  At visible wavelengths the Galactic census of
\hii\ regions thus ends at the local Solar neighborhood.  Only a
census based on observations at longer wavelengths, where the ISM is
optically thin, can locate \hii\ regions on a Galactic scale.

\hii\ regions emit at radio wavelengths because of their thermal
free-free continuum (Bremsstrahlung) and radio recombination line
emission.  Radio continuum observations of the Milky Way began in the
1950s \citep[see][]{westerhout58, piddington51, hanbury53, scheuer53,
  haddock54b}.  Optically visible \hii\ regions were also bright at
radio wavelengths.  Moreover, most of the bright, discrete radio
continuum sources in the Galactic plane turned out to be thermally
emitting \hii\ regions.
For example, of the 74 continuum sources in the \citet{westerhaut58}
catalog (not counting W82, the Moon, and W24, the Galactic center), 55
(75\%) are thermally emitting \hii\ regions, 11 are non-thermally
emitting supernova remnants (SNRs), and 7 are non-thermally emitting
extra-galactic active galactic nuclei (AGN).  Of the 55 Westerhout
\hii\ region sources, 30 are found in the catalogs of optical nebulae
\citep{sharpless53, sharpless59, rcw}.

Radio recombination lines from Galactic \hii\ regions were discovered
by \citet{hoglund65} who detected \hal{109} emission from M\,17 and
Orion\,A.  Because the Galactic ISM is optically thin at centimeter
wavelengths, RRL surveys were able to discover large numbers of \hii
regions distributed throughout the entire Galactic disk. These
pioneering surveys were done in the 1960s, 1970s, and 1980s by
\citet[][the latter hereafter referred to as L89] {mezger67, wilson70,
  reifenstein70, downes80, caswell87, lockman89}.  These surveys gave
important insights into Galactic structure and the spatial
distribution of massive star formation.  Particularly noteworthy was
the discovery of a metallicity gradient across the Galactic disk, made
apparent by RRL measurements of \hii region electron temperatures
\citep{wink83, shaver83, quireza06b}.  By the time of the L89 survey,
however, almost all of the reasonably strong radio continuum sources
had been observed; the \citet{lockman96} study of ``diffuse''
\hii\ regions was the last large angular scale survey for discrete
\hii regions made using \cm-wavelength RRLs as tracers.

Most recent work on Galactic \hii\ regions has focused on the
ultra-compact (UC) class of nebulae, which nominally are in an early
phase of \hii\ region evolution.  \citet{wc89b} and \citet{hughes89}
used the colors (flux ratios) of sources from the {\it Infrared
  Astronomical Satellite} ({\it IRAS}) point source catalog
\citep{beichman88} to identify candidate UC \hii\ regions.
\hii\ regions are bright at infrared (IR) wavelengths largely because
radiation from the central star (or stars) is absorbed by local dust,
which thermally re-emits at IR wavelengths.  Later surveys of the radio
continuum \citep{wc89a, kurtz94} and RRL emission \citep{araya02,
  watson03, sewilo04} of these targets showed that IRAS color criteria
can be used to identify UC \hii\ regions.

Despite these efforts, the census of Galactic \hii\ regions was
clearly incomplete.  The advent of modern high-resolution,
Galactic-scale infrared and radio surveys (see \S \ref{sec:sample}),
coupled with the unprecedented spectral sensitivity of the NRAO Green
Bank Telescope (GBT), has made the GBT \hii Region Discovery Survey
(HRDS) possible \citep{bania10}.  The new radio and IR surveys have
greatly increased our ability to identify Galactic \hii\ region
candidates \citep{giveon05, giveon08}.  The increases in sensitivity
due to advances in instrumentation and telescope construction allow us
to detect \hii\ region candidates in radio continuum and RRL emission.
In particular, the GBT's unblocked aperture, active surface, and
AutoCorrelation Spectrometer give it unprecedented spectral
sensitivity at centimeter-wavelengths.

The GBT HRDS extends over 168 square degrees, covering the zone from
$343\,\arcdeg \le\ \gl \le\ 67\,\arcdeg$ with $\absb \le 1\,\arcdeg$.
It detected 603 discrete hydrogen RRL components at 9\ghz (3\cm;
``X-band'') from 448 targets \citep{bania10}.  Here we provide a
catalog of the measured properties of the RRL and continuum emission
from the GBT HRDS nebulae, compare these properties to the previously
known census of Galactic \hii\ regions, and identify astrophysically
important classes of sources discovered by the HRDS.  These include
populations of \hii\ regions that lie at extreme distances from the
Sun in the Outer Arm and that are associated with {\it Spitzer}
GLIMPSE 8.0\,\micron\ ``bubble'' sources.

\section{The HRDS Target Sample\label{sec:sample}}

We assembled our target list from the following multi-frequency, large
solid angle Galactic surveys:  
the NRAO VLA Galactic Plane Survey at 21\cm\,\hi and continuum
\citep[VGPS:][]{stil06}, 
the NRAO VLA Sky Survey at 20\cm continuum \citep[NVSS:][]{condon98},  
the Southern Galactic Plane Survey at 21\cm\,\hi and continuum 
\citep[SGPS:][]{haverkorn06}, 
the VLA MAGPIS at 20\cm continuum \citep{helfand06}, and 
the {\em Spitzer} 24\micron\ MIPSGAL survey \citep{carey09}.

The HRDS candidate \hii\ regions are targets that: (1) have spatially
coincident 24\,\micron\ MIR and 20\cm radio continuum emission of a
similar morphology and angular extent, (2) have an extrapolated 9\,GHz
flux density of at least 70\,mJy, (3) are not known to be an
\hii\ region because of a previous RRL detection, and (4) are not
known to be a planetary nebula (PN), supernova remnant (SNR), pulsar
wind nebula (PWN), Active Galactic Nucleus (AGN), or a luminous blue
variable star (LBV).  Sources having both MIR and radio continuum
emission are likely emitting thermally \citep{haslam87, broadbent89},
and are therefore probably either \hii\ regions or planetary nebulae.
The plasma ionized by the FUV radiation from the exciting star(s)
gives rise to free-free thermal emission at \cm-wavelengths.  Warm
dust in the nebula absorbs the stellar radiation as well and re-emits
in the MIR.  Furthermore, small dust grains, which are stochastically
heated by the stellar flux, also emit in the MIR.


For the radio continuum  we primarily use data from the VGPS.
This survey measured 21\,cm\ \hi\ line emission, but produced
21\,cm\ continuum maps from spectral channels with no line emission.
The VGPS extends from $17\,\fdg5 \le \gl \le 67\,\fdg5$ with latitude
coverage varying between $1\,\fdg3$ and $2\,\fdg6$ in \absb and it has
an angular resolution of $1\arcmin$.  The VGPS combines data from the
GBT to fill in the zero-spacing information and is therefore sensitive
to both small and large-scale emission features.
For the longitude range $358\arcdeg \le \gl \le 17\,\fdg5$, we use the
NRAO NVSS 20\,cm\ continuum data.  The NVSS has an angular resolution
of $45\arcsec$ and covers the entire sky north of $-40\arcdeg$
Declination.  It does not include zero-spacing information and
therefore some larger, diffuse emission regions are not detected.
For $343\arcdeg \le \gl \le 358\arcdeg$, we use the continuum data
from the SGPS, which is the southern counterpart to the VGPS.  The
data were obtained in a similar manner using ATCA for the 21\,cm
interferometry measurements and the Parkes telescope for the
zero-spacing information.  The resolution of the SGPS is $100\arcsec$
and it extends from $253\arcdeg \le \gl \le 358\arcdeg$ with $\absb
\le 1\arcdeg$.
Finally, to get the best positions for targets in regions of complex
continuum emission, we use the MAGPIS 20\,cm data.  MAGPIS was made
with the VLA and extends from $5\arcdeg \le \gl \le 48\arcdeg$ with
$\absb \le 0\,\fdg8$ at an angular resolution of $\sim\,5 \arcsec$.
Although MAGPIS has the zero-spacing information from the Effelsberg
100\m telescope, visual inspection shows that diffuse emission
detected with the VGPS is often not seen by MAGPIS.  We prefer to use
the VGPS, NVSS, and SGPS since their resolutions are more comparable
to that of the GBT at X-band, and only use MAGPIS to disentangle complex
emission regions.

For the MIR emission we use MIPSGAL 24\,\micron\ data, which were
obtained with the {\it Multiband Imaging Photometer for Spitzer}
\citep[MIPS:][]{rieke04} on the {\it Spitzer Space Telescope}.
MIPSGAL covers $293\arcdeg \le \gl \le 67\arcdeg$ with $\absb \le
1\arcdeg$ at a resolution of $6\,\arcsec$ at 24\,\micron.  Due to the
low 24\,\micron\ optical depth of the intervening ISM, MIPSGAL can
detect \hii\ regions on the far side of the Galaxy.  \hii\ regions are
bright in the 24\,\micron\ MIPSGAL band for two reasons.  First, there
is thermal emission from dust grains spatially coincident with the
GLIMPSE 8.0\,\micron\ emission (see below) produced by the
\hii\ region photo-dissociation region (PDR).  There is also emission
that is spatially coincident with the ionized gas.  This emission is
most likely produced by small dust grains that are stochastically
heated by absorbing ultraviolet photons from the exciting star(s).
The grains have a temperature of $\gtrsim100$\,\K and they are not in
thermal equilibrium.  At 24\,\micron\ the flux of both components is
roughly equal for \hii\ regions surrounded by 8.0\,\micron\ bubbles
\citep{deharveng10}.

Our analysis here also uses 8.0\,\micron\ data from the Galactic
Legacy Infrared Mid-Plane Survey Extraordinaire
\citep[GLIMPSE:][]{benjamin03}, which were obtained with the {\it Infrared
  Array Camera} \citep[IRAC:][]{fazio04} on the {\it Spitzer Space
  Telescope}.  GLIMPSE has the same Galactic coverage as MIPSGAL, but
with a resolution of $\sim2\,\arcsec$ at 8.0\,\micron.  In addition to
the re-radiated emission from heated dust grains, the IRAC
8.0\,\micron\ band contains emission from polycyclic aromatic
hydrocarbons (PAHs).  These molecules fluoresce when they absorb
ultraviolet radiation, and are thus an excellent tracer of
\hii\ region PDRs.

\noindent{\it Criterion 1: Spatially Coincident 20\,cm and 24\,\micron\ Emission}
To find \hii\ region candidate targets we seek sources with matching
20\,\cm and 24\,\micron\ emission.  The MIR flux associated with
\hii\ regions is orders of magnitude greater than the radio continuum
flux \citep[e.g.][]{wc89a}.  Since the MIPSGAL $5~\sigma$ sensitivity
is 1.7\,mJy \citep{carey08}, we expect that all \hii\ regions with
9\,GHz fluxes $\gsim\,70$\,mJy (see Criterion~2 below) should be
  easily detected by MIPSGAL.

To find our \hii\ region candidates we visually compare the 20\,\cm
radio continuum emission at a particular target location with the
24\,\micron\ emission at the corresponding position, searching for
sources with spatially coincident radio and MIR emission that also
have a similar angular extent.  We did not automate our method because
of the wide range of \hii\ region morphologies, together with the
complex environments in which \hii\ regions are found.  Automated
detection methods would be extremely difficult to implement, and would
likely fail in complex regions.  Visual inspection ensures that the
target catalog includes both small, compact sources as well as larger,
diffuse regions that automated detection might confuse with the
background.  To search for \hii\ regions, we use the DS9
software\footnote{http://hea-www.harvard.edu/RD/ds9/} \citep{ds9} to
align the radio and MIR images so that their astrometry matches.  We
then repeatedly blink between these images in order to find the
nominal position of the target, which is the visually determined peak
of the radio emission.  In order to find all possible targets, we made
this analysis for each field at least three times.  Each pass through
the data added new targets.  By the third pass only a small number
were added.  These were mostly faint sources, but some came from
complex fields.  After the list was complete, we made a final pass
to refine the nominal target positions.
In addition to this ``blind'' search, we also inspected by blinking the
positions of sources found in molecular CS~($2\rightarrow1$)
\citep{bronfman96} and IR ``bubble'' \citep{churchwell06,
  churchwell07} catalogs.  These objects are prime 
\hii\ region candidates.

\noindent{\it Criterion 2: Flux Density}
The candidate targets must have an extrapolated 9\ghz flux
density greater than 70\mjy.  This is the flux of the thermal
continuum radiation from \hii\ regions ionized by a single O-9 star at
a distance of 20\kpc (see \S \ref{sec:spectral_type}).  There is no
extant continuum survey at 9\ghz with the sky coverage of the HRDS so
we must get the fluxes of target \hii\ regions by extrapolating source
fluxes from radio continuum surveys made at other wavelengths.  We
assume that the sources are optically thin at 20\cm wavelength, so
$S_\nu \propto \nu^{-0.1}$.  We use our {\it Kang}
software\footnote{http://www.bu.edu/iar/kang/} to estimate the 9\ghz
continuum flux of each candidate.  Our {\it Kang} photometry places a
circular aperture on the source and an annulus around that zone for
the background.  To be detectable in our survey, optically thin
20\cm\ \hii\ region candidates must then have VGPS or NVSS flux
densities $\gsim\,70\mjy \times(1.4\ghz/9\ghz)^{-0.1}\approx85\mjy$.
We delete all sources not meeting this flux threshold from the target
list.

\noindent{\it Criteria 3 \& 4:  Removing Previously Known H\,{\small\it II}
Regions and Other Objects}
Our \hii\ region candidate criteria can be met by a number of
astrophysical objects other than \hii\ regions.  Active galactic
nuclei (AGN), supernova remnants (SNRs), pulsar wind nebulae (PWNe),
luminous blue variables (LBVs), and planetary nebulae (PNe) can all
contaminate the target sample.  Using the SIMBAD
database\footnote{http://simbad.u-strasbg.fr/simbad/.} we remove such 
contaminating sources by searching for previously known objects that
are located within $5\,\arcmin$ of the nominal position of all the
HRDS targets.  By correlating the candidate sources with the SIMBAD
database in this way, we also remove known \hii\ regions with measured
RRL emission.

It is certainly true that the radio continuum can result from a
mixture of free-free (thermal) and synchrotron (non-thermal) emission.
Non-thermal emitters such as AGNs, SNRs, and PWNe, however, cannot
contribute many contaminating sources to the sample.  \citet{furst87}
showed that one can discriminate thermally from non-thermally emitting
objects by using the ratio of the infrared to radio fluxes.  The
IR/radio flux ratio for \hii regions is typically $\sim$\,100 times
larger than that for non-thermally emitting SNRs, so it is easy to
differentiate between the two by eye.  Our visual inspection of the
MIR and radio images should eliminate all the SNRs from our target
sample. Furthermore, since non-thermal emitting sources do not emit
RRLs, even if they were on the target list they would not appear in
the HRDS catalog of detections (see \S\ref{sec:nondetections}).
Sources showing coincident MIR and \cm-wave continuum emission
almost invariably are thermally emitting: 95\% of our sample targets
show hydrogen RRL emission with line to continuum ratios of
$\sim$\,\expo{-1} which together suggests that our targets are
emitting thermally.


Since PNe emit thermally, they are by far the largest source of
contamination.  Correlation with SIMBAD helps remove previously known
PNe, but there might be many uncataloged PNe that do not appear in
SIMBAD.  PNe do, however, have broader lines than \hii\ regions
because of their expansion \citep[see][]{garay89, balser97}, which in
principle allows the two populations to be separated ex post facto.
Based on their line widths and compact IR emission, a small number of
the HRDS sources are probably uncataloged PNe (see
\S\ref{sec:indiv_sources}).

Because of the emission from their ejected material, LBV and other
evolved stars with ejected nebulae may also appear bright at MIR and
radio wavelengths \citep[see][for their MIR emission
characteristics]{stephenson92, clark03, mizuno10}.  Most LBVs are
removed by correlation with the SIMBAD database using the nominal HRDS
target positions.  We also identify, and then remove, LBVs by visual
examination, since they appear as nearly circular objects in MIPSGAL
with very thin emission rings, whereas \hii\ regions generally show
less symmetry and broader PDRs.

\begin{figure} \centering
\includegraphics[scale=0.80]{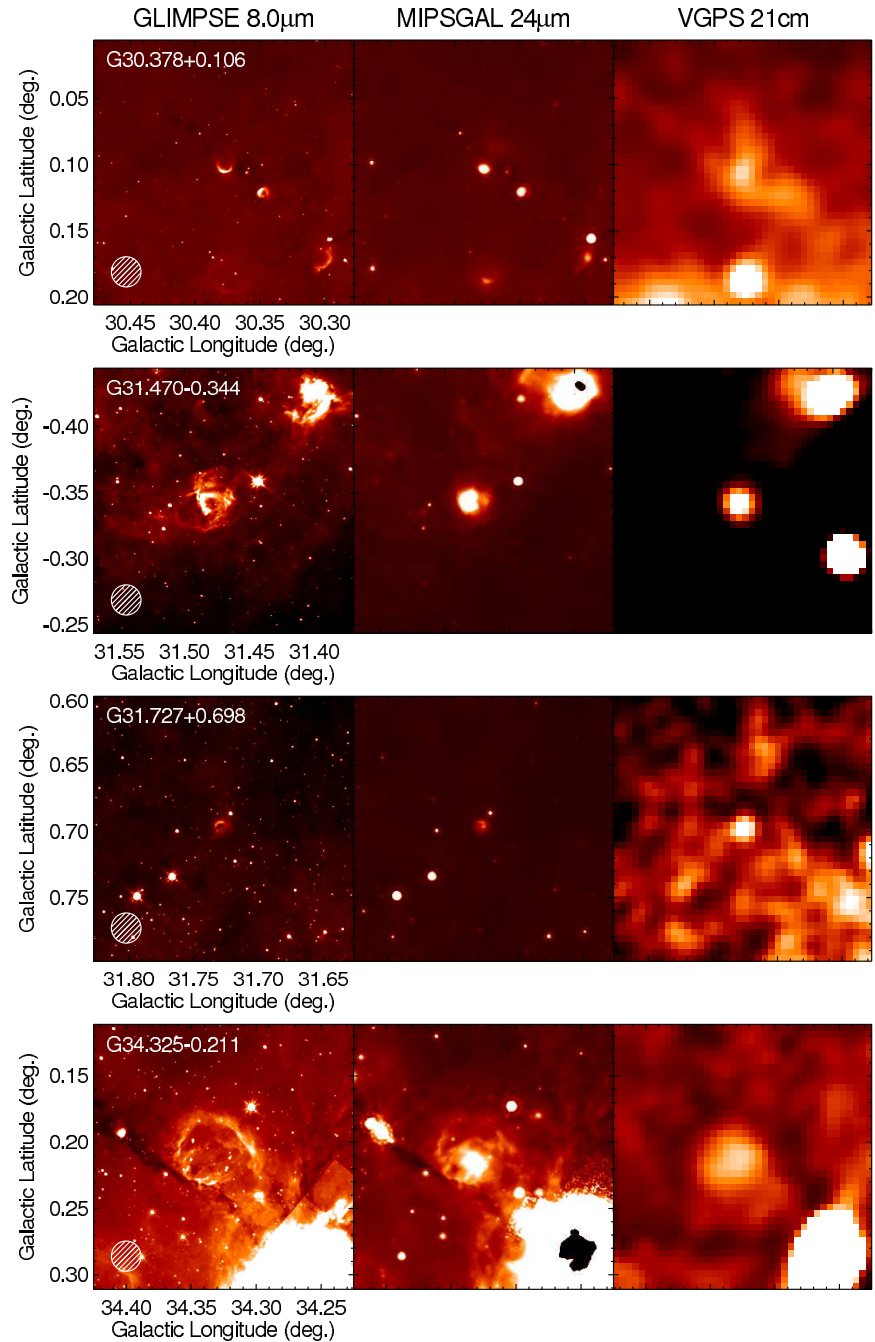}
\caption{Multi-wavelength infrared and radio images of four HRDS targets.  
From left to right are images made from Spitzer GLIMPSE 8.0\,\micron, 
Spitzer MIPSGAL 24\,\micron, and NRAO VGPS 21\,cm continuum data.  
The HRDS target is at the center of each image.  All panels are 
$5\arcmin$ on a side.  The 82\arcsec\ GBT resolution (FWHM beam size) 
is shown as a hatched circle.}

\label{fig:survey_examples}
\end{figure}

In the HRDS survey zone we find over 1,000 \hii\ region candidates
that have spatially coincident MIR and radio continuum emission with a
similar morphology and angular extent, criterion (1).  The flux cutoff
of 70\,mJy at 9\,GHz, criterion (2), reduces this number to $\sim 600$
sources, and removing known \hii\ regions and contaminating sources,
criteria (3) and (4), gives a final \hii\ region candidate list of 470
targets.  Images of the MIR and radio continuum emission for four
typical HRDS targets are shown in Figure~\ref{fig:survey_examples}.
These images are $5\arcmin$ on a side and show GLIMPSE
8.0\,\micron\ (left column), MIPSGAL 24\,\micron\ (middle column), and
VGPS 21\,cm (right column) emission for each target.  Evident is the
similar angular extent of these targets.  The environment surrounding
the targets can be quite complex.  The ``bubble'' morphology seen in
the GLIMPSE images is a common feature of our targets (see \S
\ref{sec:bubbles}).

\section{Observations and Data Reduction\label{sec:data}}

Our observations were made with the GBT 100\,m telescope from June
2008 through October 2010.  For each candidate \hii\ region, we
interleaved spectral line (using the AutoCorrelation Spectrometer
  [ACS]) and continuum (using the Digital Continuum Receiver [DCR])
observations.  Interleaving observations in this way gives a RRL
line-to-continuum ratio that is nearly independent of calibration
uncertainties and weather conditions.  We focused the telescope and
established local pointing corrections every few hours using standard
X-band pointing sources.

The calibration of the intensity scale was measured to be within 10\%
for the RRL and continuum data.  The intensity scale was determined by
injecting noise of a known magnitude into the signal path at a rate of
$1\,$Hz and $10\,$Hz for the RRL and continuum observations,
respectively, with a duty cycle of 50\%.  The antenna temperature for
the line and continuum data was set using the measured calibration
intensity, $T_{\rm cal}$, of the noise source.  The value of $T_{\rm
  cal}$ varies with frequency and was measured in the lab at
50\mhz\ intervals using hot and cold loads.  We confirmed that the
$T_{\rm cal}$ values are accurate with an uncertainty of less than
10\% by periodically observing the flux density calibration sources
3C\,147 and 3C\,286 under good weather conditions near transit.  The
flux densities for these calibrators are from \citet{peng00} and we
assume a telescope gain of $2\,$ K$\,$Jy$^{-1}$ \citep{ghigo01}.  We
did not make any corrections for atmospheric opacity or telescope gain
as a function of elevation.  Under most weather conditions opacity
corrections are typically a few percent and at 10\ghz any elevation
gain corrections are less than 5\% \citep{ghigo01}.  Overall, the
intensity scale in each ACS band is consistent at the $\sim\,12\%$
level.

To maximize the power and flexibility of our analysis we wrote a large
suite of IDL procedures rather than use any of the standard
single-dish radio astronomy software packages.  We use our own TMBIDL
software package
\footnote{See http://www.bu.edu/iar/tmbidl}
to analyze both the line and continuum GBT data.  Written in IDL,
TMBIDL inspired NRAO to develop their GBTIDL software\footnote{See
  http://gbtidl.sourceforge.net/}, which is now the main data analysis
tool for the GBT.  We find, however, that TMBIDL is much more flexible
for our needs.  We wrote extensive IDL code to analyze ACS and DCR
data both in real time and also for post data acquisition.  This HRDS
software led to the latest V6.1 version of the TMBIDL package (Bania
2010, private communication).

\subsection{Radio Recombination Lines}

The sensitivity of the GBT and the power of its ACS together made the
HRDS possible.  To achieve high spectral sensitivity, we used
techniques pioneered by \citet{balser06}, who realized that there are
8 \hna\ RRL transitions, \hal{86} to \hal{93}, that fall within the
instantaneous bandwidth of the GBT X-band receiver.  They can all be
measured simultaneously by the GBT with the ACS.  The \hal{86}
transition, however, is spectrally compromised by the
  H\,108\,$\beta$ line at nearly the same frequency so it cannot be
used for the HRDS. We thus get 14 independent spectra per OffOn total
power observation pair (7 transitions $\times$ 2 orthogonal
polarizations each).
We identify the strongest spectral feature in each band as an
\hna\ transition because there are no other strong atomic, molecular,
or unidentified lines known in these frequency bands.  For sources
with strong continuum emission we may also detect the helium and
carbon RRLs.

The RRL spectra were gotten by using the ACS in total-power, position
switching mode with On-- and Off--source integrations of six minutes
each (hereafter a ``pair'').  The Off--source position observation
tracked the same azimuth and zenith angle path as the On--source scan.
The ACS was configured as 16 independent 50\,MHz wide spectral bands.
Each was 9-level sampled by 4096 channels.  The typical velocity
resolution of any band is $\sim\,0.4\,\kms$ per channel.  Because the
target RRL velocities were unknown, all ACS bands were tuned to a
center LSR \footnote[5]{The RRL velocities here are in the kinematic local
standard of rest (LSR) frame using the radio definition of the
Doppler shift.  The kinematic LSR is defined by the solar motion of
20.0\kms toward \radec = $(18^h, +30\degree)$ [1900.0] \citep{gordon76}.}  velocity
of 42.5\,\kms\ for targets in the \gl range from $10\,\arcdeg$ to
$67\,\arcdeg$.  For targets in the \gl = $343\,\arcdeg$ to
$10\,\arcdeg$ range the center velocity was 0.0\,\kms. The 50\,MHz ACS
bands span $\sim\,1,700\,\kms$ at X-band, which greatly exceeds the
LSR velocity range of objects gravitationally bound to the Milky Way.
We can thus, in principle, detect RRL emission from any Galactic
\hii\ region.

\citet{balser06} showed that all these transitions can be averaged
together to improve the RRL signal-to-noise ratio, thus giving an
extremely sensitive X-band \hnaa\ composite nebular spectrum.  Because
the energy level spacings between adjacent RRLs for ${\rm n} > 50$ are
similar relative to the ground state, the line parameters ---
intensity, line width, and velocity --- are nearly identical.  For
example, the classical oscillator strengths for the \hal{50} and
\hal{51} RRLs differ by only 2\%, so these lines should have
comparable intensities (Menzel 1968).  Thus at the high energy levels
of our observations, to first order all transitions have the same
intensity.

We can thus average all the 14 independent spectra together and get an
extremely sensitive \hnaa\ composite RRL spectrum for a source.  
Three steps are required to average the individual \hna\ RRLs into a
source composite \hnaa\ spectrum:
(1) the velocity scale must be re-gridded to a common resolution;
(2) the RRL spectra must be shifted to match the Doppler tracked \hal{89} RRL; and
(3) the spectra must be aligned in velocity and averaged together.
The first step is required because the velocity resolution is a
function of sky frequency, which varies with the different RRL
transition frequencies.  We therefore re-grid the \hal{88} to \hal{93}
RRL spectra to the velocity scale of the \hal{87} spectrum, which has
the poorest spectral resolution.  The second step is necessary since
in the GBT system only the \hal{89} spectral band is properly Doppler
tracked.  For the third step, we align the velocities and the average
spectra in the standard way with a weighting factor of $t_{\rm
  intg}/T_{\rm sys}^2$, where $t_{\rm intg}$ is the integration time
and $T_{\rm sys}$ is the total system temperature.

We smooth the \hnaa\ spectrum with a normalized Gaussian over 5
channels to give a spectrum with a velocity resolution of 1.86\,\kms,
which compares favorably with the $\sim\,25$\,\kms\ FWHM line widths
that are typical for RRLs from Galactic \hii\ regions.  We then remove
a polynomial baseline, which is usually a second order fit.  Finally,
we fit Gaussians to this final composite \hnaa\ spectrum, which gives
us the LSR velocity, line intensity and line width for each emission
component.  The baseline removal and Gaussian fitting procedures are
not automated, but rather are done individually for each source.

This observing technique, coupled with the sensitivity afforded by the
GBT's aperture, gave us unprecedented spectral sensitivity per unit
observing time advantage compared with all previous \cm-wavelength RRL
surveys of Galactic \hii regions.  The vast majority of our targets
took only a single OffOn total power pair to detect RRL emission and
establish an accurate source LSR velocity.  Typically, the r.m.s. noise
for observations consisting of a single pair is $\sim 1$\,mJy.  For
some of the weaker sources, and for sources for which we wanted to
measure the H/He line ratio more accurately, we made additional
observations.

Over the frequency span of the \hal{87} to \hal{93} RRLs, the GBT HPBW
beam size varies by 20\%.  We therefore expect that source structure
will make the various transition intensities differ by more than what
is theoretically expected.  We observed this effect for bright,
compact \hii\ regions that we used as calibrators, such as W3, where
the RRL intensities approximately scale with the beam area, as is
expected for a point source.  For the HRDS catalog, we did not scale
to a common beam size; the \hnaa\ intensities are averages over the
beam size range of $73\,\arcsec$ to $89\,\arcsec$.

We show in Figure~\ref{fig:example_lines} \hnaa\ spectra for the
Figure~\ref{fig:survey_examples} \hii\ regions.  These are typical
HRDS \hnaa\ spectra.  The G\,31.727+0.698 nebula, however, is one of
our weakest detections.  Both hydrogen and helium RRLs can be seen in
the G\,31.470$-$0.344 spectrum.  We detect two RRL components along
the line of sight to G\,30.378+0.106, one likely due to W43 (see
\S\ref{sec:multiples}).

\subsection{Radio Continuum\label{sec:radio_continuum}}

We used the DCR to measure the continuum flux for our HRDS targets by
slewing the telescope across the sky while sampling the total power
every 100\,ms.  We made cross scans centered on the nominal source
position consisting of four total-power integrations: forward and
backward in Right Ascension, RA, and forward and backward in
Declination, Dec.  Each scan was $20\,\arcmin$ in length and we slewed
at $20\,\arcsec$ per second.  The DCR was centered at 8665 MHz with a
320 MHz bandwidth.

The continuum data can be considerably more complex than the RRL data
due to a combination of source confusion and sky noise.  A typical
observation of an HRDS target starts with a continuum measurement
followed by a spectral line pair.  For some sources, however,  we have
many such continuum scans.  We thus average all the forward and backward
scans (after first flipping the backwards scans).  This average
usually blends different observing epochs and weather conditions
so it tends to minimize the effects of sky noise.  We remove a
polynomial baseline, usually a second order fit, from the RA and
Dec average scans and then fit Gaussians to them.  This analysis 
cannot be automated and was done on a source by source basis. 

The inner Galaxy has a high density of sources that are not always
well-separated by the $82\,\arcsec$ GBT beam.  We attempt to resolve
this confusion by visually examining VGPS, NVSS, or SGPS images.
These data helped us to separate the target's emission from that of
other contaminating sources and sky noise.  Each source was analyzed
individually and the polynomial baseline was removed based on what we
judged was the baseline zero intensity level of the scan.  Some
sources showed multiple Gaussian components.  Sometimes all
components can be associated with emission from the HRDS target.  In
other cases one or more Gaussian components are due to other,
unassociated radio continuum sources.  The auxiliary radio continuum
data from the various sky surveys provided vital extra information
that we used to decide which components are due to the HRDS target.

Clearly the resolution of source confusion is a subjective process.
Despite our care, the fits to the radio continuum data are, especially
for weak sources, uncertain and in some cases no fit is possible.
Figure~\ref{fig:example_continuum}, which shows the continuum scans
for the four Figure~\ref{fig:survey_examples} nebulae, illustrates the
complex nature of these data.  The thick lines show the Gaussian fit
to the HRDS component of each scan.  For sources with simple emission
structure which is also well-centered in the GBT beam, we expect the
fitted heights in the RA and Dec scans to be of equal intensity.  
Additionally, all sources should have fitted sizes at least as large
as the GBT beam.  When these conditions are not met, it can be a sign
of an inaccurate position or poor data quality.  Continuum parameters
for such sources should be used with extreme caution.


For each source we analyze the RA and Dec continuum emission scans
independently.  For targets with simple structure we fit a single
Gaussian component, which gives a peak intensity (in \mK) and full
width at half maximum (FWHM) angular size (in arcsec) for each RA and
Dec scan.  For many of the HRDS sources, however, the continuum
emission is too complex to be modeled by a single Gaussian.
For these cases, we fit the continuum scan with multiple Gaussian
components, each with its own peak intensity and FWHM angular size.
The sum of these components gives an estimate for the total source
integrated intensity (units are [\mK\,arcsec]).  We then crudely
approximate this complex emission with a single Gaussian model whose
peak intensity and FWHM angular size together produce the same
integrated intensity as the sum of the multiple components.  The FWHM
angular size of this fictitious single component is set by the maximum
separation of the FWHM position extrema of the multiple components
with respect to the nominal target position. This angular size,
together with the integrated intensity determined by the multiple
components, defines the peak intensity of the fictitious single
component.

We use the Gaussian components to estimate the source integrated flux
density at 8665\,\mhz.  We take the arithmetic average of the RA and
Dec intensities to define an average peak intensity and
use the 2\,K per Jy GBT X-band gain to convert this into the peak
flux.  Assuming the source continuum emission is a two-dimensional
Gaussian in RA and Dec, the integrated source flux, $S_i$, is:
\begin{equation}
S_i = S_p \, \left(\frac{\theta_{\rm RA}}{\theta_b}\right)
\left(\frac{\theta_{\rm Dec}}{\theta_b}\right)\,,
\label{eq:integrated_flux}
\end{equation}
where $S_p$ is the peak flux, $\theta_b$ is the GBT beam size at
X-band, and $\theta_{\rm RA}$, $\theta_{\rm Dec}$ are the FWHM angular
sizes derived from the RA and Dec scans \citep[see][]{kuchar97}.  


\section{The H\,{\bf\footnotesize II} Region Discovery Survey Catalog
\label{sec:catalog}}

We detect radio recombination line and continuum emission from 448
of our 470 targets, over 95\% of our sample of \hii\ region
candidates.  The HRDS targets with detected hydrogen RRLs are listed
in Table~\ref{tab:sources}, which gives their Galactic and Equatorial
(J2000) coordinates.  For brevity, we provide here only data for the
first 10 sources in the HRDS catalog.  These stubs show the
information given in each catalog table; the complete HRDS catalog
data are provided in the electronic version of this paper.  Table~ \ref{tab:sources}
also provides information about a source's
association with methanol masers \citep[e.g.,][]{pestalozzi05} and with
CS~$2\rightarrow 1$ emission \citep{bronfman96}, together with a
characterization of the source morphology as seen in {\it Spitzer}
GLIMPSE 8.0\,\micron\ images.
Methanol maser and CS emission are both produced in dense gas regions.
They are associated with the earliest phases of massive star
formation; these data therefore give some indication of the
evolutionary state of HRDS targets.  The method we use to associate
the maser and CS positions with the HRDS targets, as well as the
implications of these associations for the HRDS targets are discussed
in \S \ref{sec:correlation}.  Sources observed but not detected in CS
are listed as ``ND''.  The Table~\ref{tab:sources} maser velocity is
that of the emission component with the highest intensity.

Images of some example HRDS targets are shown in
Figure~\ref{fig:threecolor}.  They are made from {\it Spitzer} survey
data: MIPSGAL 24\,\micron\ (red), GLIMPSE 8.0\,\micron\ (green), and
GLIMPSE 3.6\,\micron\ (blue).  HRDS targets are often found in complex
IR environments and they have a wide range of sizes and morphologies.
Figure~\ref{fig:threecolor} shows examples of all the different
morphological classifications found in Table \ref{tab:sources}.
For nearly all the angularly resolved HRDS \hii\ regions, the
24\,\micron\ emission lies interior to the 8.0\,\micron\ emission.
This appears to be the MIR signature of Galactic \hii\ regions.
Three-color {\it Spitzer} images such as Figure~\ref{fig:threecolor}
can thus be used to find \hii\ region candidates with high
reliability.  The most striking examples of this phenomenon are the
GLIMPSE ``bubbles'' (see \S\ref{sec:bubbles}).

We detected 603 discrete hydrogen RRL components from the 448 HRDS
targets.  The RRL data are given in the Table \ref{tab:line} stub,
which lists the source name, the Galactic longitude and latitude, the
line intensity, the FWHM line width, the LSR velocity, and the r.m.s.
noise in the spectrum.  The errors given in Table \ref{tab:line} for
the fitted line parameters are the $1\,\sigma$ uncertainties from the
Gaussian fitting.  For sources with multiple velocity components
detected along the line of sight, the source names are given
additional letters, ``a'', ``b'', ``c'', or ``d'', in order of
decreasing peak line intensity.

A stub of the HRDS radio continuum catalog data is given in Table
\ref{tab:continuum}, which lists the source name, the Galactic
longitude and latitude, the peak intensity in the RA and Dec scan
directions, the FWHM angular size in the RA and Dec directions, and
the integrated flux density derived using
Equation~\ref{eq:integrated_flux}.  For sources better approximated by
multiple Gaussians, the angular size listed is the maximum extent of
the multi-component composite source and the intensity is that for a
single Gaussian component, which is determined by this width and the
integrated flux density (see \S\,\ref{sec:radio_continuum}).  Such
``complex'' objects are identified by a ``C'' label in the Notes column of Table
\ref{tab:continuum}.  The errors given for the continuum parameters are
the $1\,\sigma$ uncertainties in the Gaussian fits, propagated as
necessary for derived quantities using
Equation~\ref{eq:integrated_flux}.

The main goal of the HRDS was to detect new Galactic \hii\ regions.
To maximize the number of observed objects we therefore used standard
observing procedures throughout the survey and did not customize the
observing for specific sources.  For example, the peak continuum
emission observed with the GBT was typically offset relative to the
nominal target position, but we did not refine these coordinates,
i.e., we did not peak-up on the source relative to the GBT's beam.
Moreover, for some targets it was difficult to establish the continuum
zero-level given the complexity of the continuum emission structure on
the sky and time variable sky noise.  Source confusion or cases where
a very weak target is quite near to a strong source can compromise the
accuracy of the continuum measurement.  In some sources longer
continuum cross scans might have yielded more accurate baselines, but
our goal was to produce a new catalog of Galactic \hii\ regions not to
provide the highest quality continuum data.  Much of the continuum
data in the HRDS catalog is therefore not suitable for deriving very
accurate physical properties of these \hii\ regions, such as the
electron density and temperature, the excitation, etc.  A subset of
sources, however, have single component continuum peaks that lie
within 10\,$\arcsec$ of the nominal target position.  These nebulae, 
which are flagged with a ``P'' label in the Notes column of Table 
\ref{tab:continuum}, are our highest quality continuum data. 
They {\em are\/} suitable for deriving physical properties.  

The HRDS catalog, however, is an excellent resource for Galactic
structure studies since the nebular positions and LSR velocities are
measured with high accuracy.
Each RRL detection yields the nebular source velocity.  The nebular
velocity, when combined with the \lb\ location, gives the
source Galactocentric position and kinematic distance from the Sun.
The continuum measurements of the thermal free-free emission give the
nebular total flux, angular size, and, crudely, some knowledge of the
source morphology.  

\subsection{Sources Not Detected\label{sec:nondetections}}

We did not detect RRL emission from 22 of the candidate \hii\ region
targets.  These are typically $\lsim\,1\,\mjy$ upper limits, which is
the sensitivity of an \hnaa\ composite spectrum made from a single
observing pair. These non-detections are listed in Table
\ref{tab:nondetections}, which gives the source name, the GLIMPSE
8.0\,\micron\ morphology (using the Table \ref{tab:sources}
classifications), comments on each source, and our best estimate of
the object's type: \hii\ region, PN, AGN, or PWN.
Nearly half of these non-detections, 10 of the 22 targets, may be PNe
or emission line stars.  They are unresolved in the IR and the radio
continuum surveys we used to compile our target list.  


Five sources, 23\% of the sample of non-detections, have spatially
coincident MIR and radio emission, with morphologies that are similar
to HRDS \hii\ regions. Their extrapolated X-band fluxes made them HRDS
candidate targets. Their measured X-band continuum emission is quite
weak, however, so they are probably \hii\ regions that fall below our
detection threshold.  One source, G005.756$-$0.932\hide{FA094}, is an
IR bubble with coincident NVSS emission.  It is not included in the
\citet{churchwell07} catalog.  This source is odd since it has the
same bubble morphology at 8.0\,\micron\ and 24\,\micron, whereas all
the other \hii\ regions in our sample show 8.0\,\micron\ emission that
surrounds, and extends beyond, the 24\,\micron\ emission
\citep{deharveng10}.

The source G045.797$-$0.347\hide{LA263} is probably a region of thermally
emitting plasma too faint to be detected by the HRDS.  It may be 
a star-forming region that is unrelated to the compact sources seen
nearby in MIR images.
The G017.223+0.397\hide{LA919} target is near a \citet{lockman96}
``diffuse'' \hii\ region source.  It is not, however, a diffuse
nebulae: \citet{stephenson92} says that it is an evolved
star, which is plausible on the basis of our non-detection.  
\citet{stephenson92} also claims that G357.387+0.011\hide{FA044} 
is a distant, luminous star.

Five other sources seem to be non-thermal emitters.  We believe four
of them are AGN because of the poor alignment of their radio and IR
emission.  If a source is unresolved at $\sim1\,\arcmin$ resolution,
then the MIR and radio emission should be spatially coincident to
within $\sim30\,\arcsec$.  This is not the case for these sources,
which suggests that they are chance alignments of AGNs with Galactic
MIR sources.
The remaining source, G054.096+0.266, \hide{LA722} is a PWN;  it is
the PSR~J1930+1852 pulsar wind nebula associated with SNR~54.1+0.3.  Bright at
24\,\micron, this source is faint at 8.0\,\micron, which suggests that
line emission may be causing the 24\,\micron\ brightness.  We observed
it by mistake.

\subsection{Multiple Velocity Component Sources\label{sec:multiples}}

Of the 448 HRDS \hii\ regions, 129 (nearly 30\%) have multiple
  RRL components in their \hnaa\ spectra at different velocities.  We
find 105 sources with two components and 23 with three components.
One source has four velocity components.  With the $82\,\arcsec$ GBT
beam at X-band, it is unlikely that we would detect multiple,
physically distinct \hii\ regions along any given line of sight.  One
of the components must surely come from the HRDS target \hii\ region.
The other component may stem from more diffuse ionized gas along the
line of sight.  This diffuse component may originate from ionizing
photons leaking into the extended PDRs produced by nearby, large
star-forming regions.

Even \hii\ regions with well-defined PDRs leak radiation beyond the
ionization front \citep{zavagno07, anderson10a}.  Moreover,
\citet{oey97} compared the observed H$\alpha$ luminosity with that
predicted from stellar models and concluded that up to 50\% of the
ionizing flux was escaping from \hii\ regions.  The complex morphology
of a typical HRDS nebula, e.g. Figure~\ref{fig:threecolor}, suggests
that within any nebula some zones are ionization bounded and others
are density bounded.

If we are to derive the kinematic distances and physical properties of
HDRS nebulae, we must unambiguously identify their velocities in these
multi-component spectra.  The resolution of this problem is beyond the
scope of this paper.  Nonetheless, here we make a preliminary analysis
of HRDS sources located near the W43 star forming \hii\ region
complex.

Over 30\% of the HRDS multiple velocity component sources are located
within $2\arcdeg$ of the large, massive star forming region W43.  This
is a mere $\sim2\%$ of the survey area; it represents a significant
overabundance of multiple velocity component sources in a very small
zone of sky.  The RRL LSR velocity of W43, measured from a single
$\sim3\,\arcmin$ resolution pointing, is 91.6\,\kms\ (L89).  This
velocity corresponds to a kinematic distance of 5.7\,kpc
\citep{anderson09a}, which places W43 at the end of the Galactic bar
\citep{benjamin05}.  From studies of external Galaxies, we know that
the end of galactic bars can have significant star formation
\citep[e.g.][]{martin97}.  The entire W43 complex has velocities
ranging from $\sim80$ to $105$\,\kms, when mapped at high angular
resolution with the VLA \citep{balser01}.

Of the 23 HRDS sources within $0\,\fdg5$ of W43, 21 have multiple
velocity components.  The two single component sources are compact
\hii\ regions that share the same velocity as W43.  All the other
sources have one velocity component in the RRL velocity range found by
\citet{balser01}.  The other component is from the HRDS target source.
The origin of the multiple-velocity components for the sources in the
direction of W43 appears to be low density ionized gas along the line
of sight.  It is at present unclear if this low-density plasma is from
FUV photons leaking into the extended PDR surrounding W43, from plasma
not necessarily associated with W43 at the end of the Galactic bar, or
from both.

\begin{figure} \centering
\includegraphics[width=6 in]{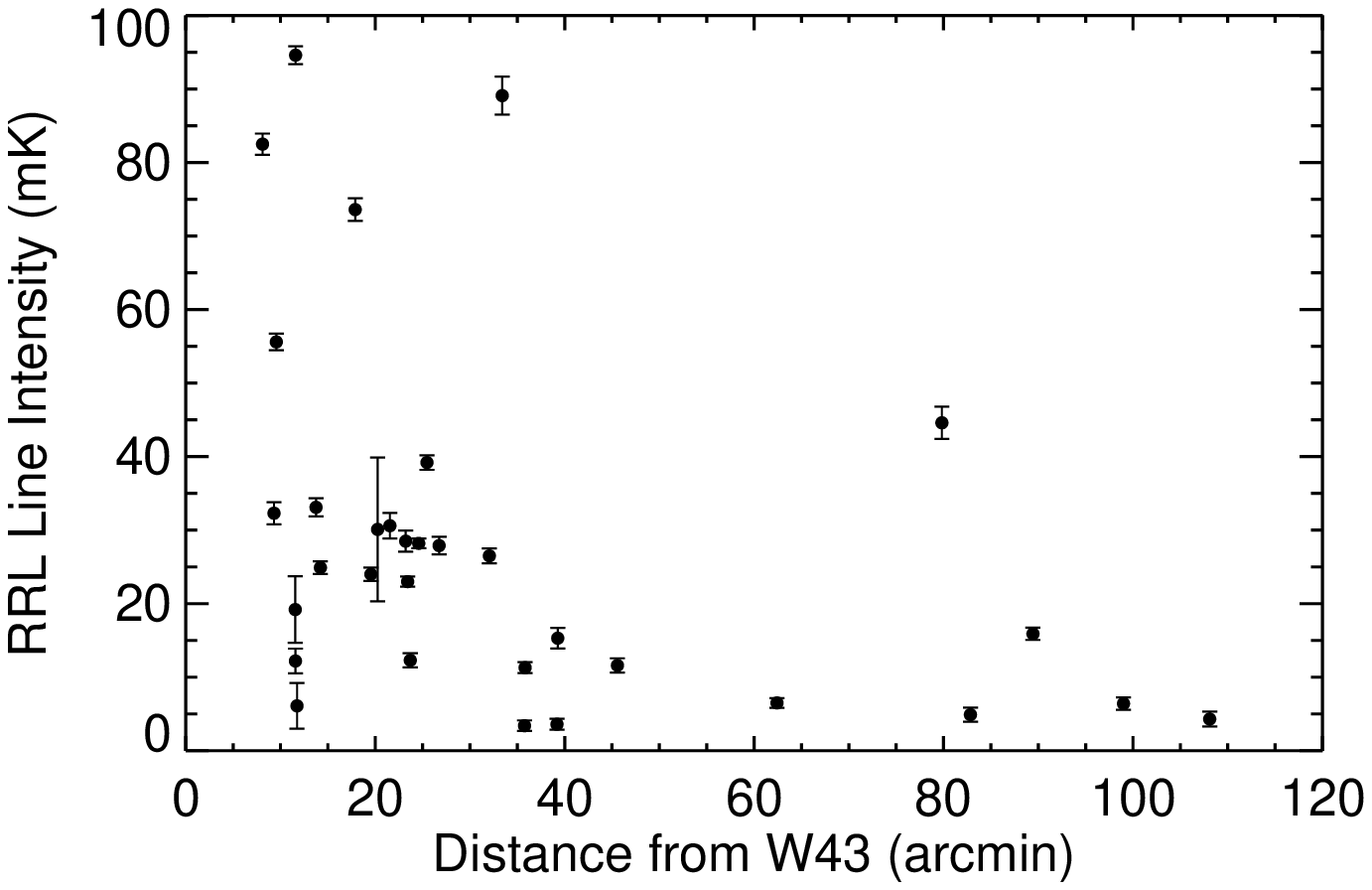}

\caption{RRL intensity as a function of angular distance from W43.
Shown is the intensity of the 34 HRDS RRL components within
$2\,\arcdeg$ of the nominal center position of W43, \lb\ =
(30\,\fdg78, $-$0\,\fdg03), whose velocities lie within the W43
velocity range, 85 to 105\,\kms \citep{balser01}, plotted as a
function of the angular distance from the nominal center position.  
Error bars show the $\pm3\,\sigma$ uncertainties in Gaussian fits to
the hydrogen RRL spectra.  The decrease in RRL intensity with distance
suggests that many of these multiple velocity components are
associated with the W43 star forming complex.}
\label{fig:w43_lines}
\end{figure}

If the origin of the multiple-velocity components is leaked plasma
from W43 itself, we would expect the intensity of a source's W43 RRL
component to decrease, on average, with its distance from the nominal
center of W43, \lb = (30\,\fdg78, $-$0\,\fdg03).  Here we are assuming
that the intensity of the RRL emission decreases from the center of
W43 outwards, just as the continuum emission does.  In
Figure~\ref{fig:w43_lines}, we plot the intensity of an HRDS source's
W43 RRL component as a function of its distance from the nominal
center of W43.  Shown is the intensity of the RRL component whose
velocity is nearest to the 91.6\,\kms velocity of W43.
Figure~\ref{fig:w43_lines} shows that these components are on average
stronger for sources near W43.  This is consistent with the hypothesis
that these spectral lines are caused by an extended, partially ionized
PDR surrounding W43.

This interpretation, however, is far from secure.  Given the distance
to W43, an area $1\degree$ in diameter would be $\sim100$\,pc across.
This is extremely large, although not unheard of, for a star forming
complex.  From thermal radio sources such as W43, when RRL emission is
detected we should also be able to detect radio continuum emission.
The lack of extended faint continuum emission associated with W43 in
the direction of these multiple-velocity RRL sources is therefore also
a problem for this interpretation.  Perhaps we are detecting a
combination of diffuse emission leaked from W43 and plasma at the end
of the Galactic bar.

It is similarly unclear what is causing the multiple-velocity
components for nearly all other multiple-velocity HRDS sources.  These
lines of sight must be individually analyzed in order to determine
which RRL velocity component is associated with the HRDS compact MIR
target.  This topic will be addressed in future papers.

%

\subsection{Infrared ``Bubble'' Sources\label{sec:bubbles}}

Many {\it Spitzer} GLIMPSE ``bubble'' sources seem to be
associated with \hii\ regions \citep{bania10, deharveng10}.  
We detected RRLs from the ionized gas in 65
GLIMPSE ``bubble'' sources cataloged by \citet{churchwell06,
churchwell07}.  
By visually examining GLIMPSE images for the entire HRDS zone,
$343\,\arcdeg \le \gl \le 67\,\arcdeg$ with 
$\absb \le 1\,\arcdeg$, we find that 170 of the 420 cataloged GLIMPSE
bubbles, $\sim\,40\%$ of the sample, are associated with previously
known \hii\ regions.  When these are combined with the 65 HRDS
detections, more than half of all cataloged MIR bubbles turn out to
be \hii\ regions with measured RRL emission.

Altogether there are 207 HRDS sources with a bubble morphology; these
are targets classified as ``B,'' ``BB,'' ``PB,'' or ``IB'' in Table
\ref{tab:sources}.  These additional 142 HRDS nebulae with a bubble
morphology that are not in \citet{churchwell06, churchwell07} imply
that there is a large population of undiscovered bubbles.  Within the
HRDS zone there are in fact hundreds more cataloged and uncataloged
GLIMPSE bubbles that have spatially coincident radio continuum and
MIPSGAL 24\,\micron\ emission.  They are all \hii\ region candidate
targets.  They are, however, also very weak continuum sources; all are
weaker than the HRDS target flux threshold at 9\,\ghz.

%

We speculate that nearly all GLIMPSE bubbles are caused by
\hii\ regions \citep{bania10}.  In the HRDS zone, the combined sample
of HRDS and previously known \hii\ regions with measured RRL emission
contains $\sim\,900$ nebulae (see \S\,\ref{sec:properties}).  More than
40\% have a bubble morphology at 8.0\,\micron.  For longitudes $\gl
\ge 10\,\arcdeg$, where source confusion is not as great as toward
the Galactic center, the percentage is $\gsim 50\%$.  The bubble
morphology is obviously common for Galactic \hii\ regions.

It appears that \hii\ regions with a bubble morphology do not
represent any particular evolutionary state.  Some bubble sources are
associated with masers and CS emission, which suggests that they are
young (see Table~\ref{tab:sources} and \S\,\ref{sec:correlation}). For
others, their radio continuum emission is faint and diffuse, implying
an evolved state.  The bubble morphology is probably a consequence of
the density structure of the surrounding ISM.  In a turbulent medium,
a complete bubble would be difficult to form because a density
gradient would favor expansion in the direction of decreased density.

On the basis of CO observations, \citet{beaumont10} argue that the
``bubble'' sources are in fact physically two-dimensional rings,
rather than limb brightened projections on the sky of optically thin
three dimensional bubble structures.  In this scenario, the rings are
produced by the bipolar flows that occur during star formation.  To
make a complete ring, a bipolar flow source must be oriented at nearly
zero inclination angle with respect to the line of sight, so the flows
of ionized gas are pointing directly toward and away from the
observer.  At other inclination angles these flows make incomplete,
bipolar rings on the sky. 

The ring scenario thus predicts two observational consequences that
can be tested with the HRDS data.  First, there should be many more
bipolar bubbles than complete bubbles, since the latter would be seen
only for particular, and rare, inclination angles.  Secondly, the RRL
line widths for bubbles should be larger than those of the
\hii\ region sample at large, since both the red- and blue-shifted
ionized gas from the bipolar flow contribute to the bubble line width.

The HRDS catalog strongly suggests that bubble sources are
three-dimensional objects.  There are far more nearly complete
bubbles, 117 morphological class ``B'' sources, than bipolar nebulae:
the HRDS has only 8 morphological class ``BB'' bipolar bubbles.
Bipolar bubbles are quite rare.  Furthermore, the average RRL line
width of bubble sources is 22\,\kms (see \S\,\ref{sec:linewidth}),
which is identical to the average for the entire HRDS sample.  Both
ring scenario predictions are contradicted by the HRDS catalog nebular
statistics.  We therefore conclude that the majority of bubble sources
are three-dimensional structures.


\subsection{Notes on Individual Sources \label{sec:indiv_sources}}

\subsubsection{Possible Planetary Nebulae \label{sec:PNeNote}}

Because PNe are physically small compared to \hii\ regions, unresolved
emission seen in GLIMPSE with no nebulosity may be a hallmark of PNe
that might be confused with \hii\ regions.  Nearly all of the
8.0\,\micron\ images of HRDS sources show nebulosity.  There are,
however, 13 HRDS RRL sources that appear point-like in the GLIMPSE
data.  These objects may be PNe; they are noted in Table
\ref{tab:sources} as being PN candidates, ``PN?''.  These sources are:
G009.741+0.842, \hide{FA110} G012.116+0.076, \hide{LA905}
G012.199$-$0.034, \hide{LA1031} G016.404$-$0.575, \hide{LA1066}
G017.364+0.519, \hide{LA1073} G018.708$-$0.126, \hide{LA507}
G022.154$-$0.153, \hide{LA541} G029.874$-$0.819, \hide{LA616}
G030.234$-$0.139, \hide{LA202} G030.532$-$0.259,
\hide{LA030}\\ G030.663$-$0.334, \hide{LA301} G046.017+0.264,
\hide{LA144} and G050.556+0.045. \hide{LA702}
Four of these sources were detected as part of the RMS survey
\citep{urquhart09} and these authors classify them as PNe.  Since all
targets classified as point sources (``PS'' in
Table~\ref{tab:sources}) are similar in this regard, it seems likely
that most of them are also PNe.

Based on their 8.0\,\micron\ morphology, four additional sources are
also likely to be PNe: G003.449$-$0.647, \hide{FA086} 
G016.228$-$0.368, \hide{LA1060}  G017.414+0.377, \hide{LA1074}  and
G026.317$-$0.01. \hide{LA458}  Two of these sources,
G017.414+0.377 \hide{LA1074} and G026.317$-$0.012, \hide{LA458}  are
observed as faint rings at 8.0\,\micron.  Their low
8.0\,\micron\ intensity and morphology separate them from the other
HRDS 8.0\,\micron\ rings.  For G026.317$-$0.012, \hide{LA458}  a central
point source is detected within the ring.  The two other sources,
G003.449$-$0.647 \hide{FA086} and G016.228$-$0.368, \hide{LA1060}  are not
detected at 8.0\,\micron.  They have a ``ring'' morphology at
24\,\micron\ and are listed in the \citet{mizuno10} catalog of MIPSGAL
ring sources.  These authors suggest that a large percentage of
sources observed as rings at 24\,\micron\ are PNe or evolved stars.

The HRDS source G031.727+0.698 \hide{LA014} is also listed in the
\citet{mizuno10} catalog.  This source has a negative LSR velocity of
$-39.2$\,\kms\ and it is therefore located in the outer Galactic disk.
Its large distance gives it a high intrinsic luminosity and a large
physical size, which means that it cannot be a PN.  The minimal
overlap between the HRDS, which contains 207 8.0\,\micron\ rings
(bubbles), and the \citet{mizuno10} catalog of 24\,\micron\ rings
implies that while 8.0\,\micron\ rings nearly always surround
\hii\ regions, the rings detected at 24\,\micron\ are associated with
different objects.

Due to their expansion and higher electron temperature, the line
widths of PNe are generally greater than that of \hii\ regions
\citep[see][]{garay89, balser97}.  \citet{balser97}, for example, find
an average hydrogen recombination line width of $44.1 \pm 5.8 \kms$ for
a sample of six PNe.  The 17 HRDS sources identified here as being
possible PNe have an average line width of 25\,\kms, which is near the
mean of the entire HRDS sample, 22\,\kms.  Three sources, however, do
have broad RRLs: G026.317$-$0.012, \hide{LA458} G029.874$-$0.819,
\hide{LA616} and G050.556+0.045 \hide{LA702} have line widths of 38.0,
40.6, and 48.8\,\kms, respectively.  Surprisingly, six HRDS sources
listed as possible PNe have narrow line widths, $<17\,\kms$:
G009.741+0.842, \hide{FA110} G016.404$-$0.575, \hide{LA1066}
G017.364+0.519, \hide{LA1073} G017.414+0.377, \hide{LA1074}
G018.708$-$0.126, \hide{LA507} and G022.154$-$0.153. \hide{LA541} More
observations are needed to better understand the nature of
these sources.




\subsubsection{Non-detections at 8.0\,\micron}

Eight sources have 24\,\micron\ emission but no detectable
8.0\,\micron\ emission: G003.449$-$0.647, \hide{FA086}
G011.573+0.340, \hide{LA1018} G016.228$-$0.368, \hide{LA1060}
G019.786+0.285, \hide{LA419} G023.513$-$0.244, \hide{LA555}
G023.849$-$0.196,\\ \hide{LA438} G029.019+0.170, \hide{LA607} and
G049.507$-$0.520. \hide{LA166}  These objects are listed in Table
\ref{tab:sources} as a non-detection, ``ND''.  That we detect RRL
emission from these targets points to a thermal origin. Two of these 
sources, G003.449$-$0.647 \hide{FA086} and G016.228$-$0.368, \hide{LA1060}
are identified above as being PNe candidates.
Finally, G049.507$-$0.520 \hide{LA166} has no associated GLIMPSE or
MIPSGAL emission but is likely associated with W51 (see below).  The
four remaining sources all have an extended, curved morphology at
24\,\micron\ and are thus unlikely to be PNe.



\subsubsection{Nuclear Disk}

We found nine sources whose velocity $<-200$\,\kms\ and Galactic
location of $\gl\,\sim\,359\,\arcdeg$ place them in
the nuclear disk of the Milky Way: G358.530+0.056\hide{FA050},
G358.552$-$0.025\hide{FA051}, G358.616$-$0.076\hide{FA053},
G358.652$-$0.078\hide{FA056}, G358.680$-$0.087\hide{FA458},
G358.694$-$0.075\hide{FA058}, G358.720+0.011\hide{FA059}, 
G358.827+0.085\hide{FA062}, and G359.159-0.038\hide{FA064}.  
Before the HRDS, only three \hii\ regions
were known to share this location and velocity range: G358.623$-$0.066
at $-212$\,\kms, G358.797+0.058 at $-206.6$\,\kms, and
G358.974$-$0.021, which has two RRL components at $-193.3$ and
$-5.4$\,\kms \citep{caswell87, lockman96}.  The location observed by
\citet{caswell87}, $\lb = (358.623,\,$-$0.066)$, lies in between the
two HRDS sources G358.652$-$0.078\hide{FA056} and
G358.616$-$0.076\hide{FA053}; it is not a distinct \hii\ region.  
These HRDS sources are distributed within a Galactic zone
$\sim30\arcmin$ in diameter, or $\sim75$\,pc assuming a distance of
8.5\,kpc.  Most of these nebulae have a similar compact morphology (we
classify five of them as ``Compact'' in Table~\ref{tab:sources})
suggestive of relatively early evolutionary stages.

\subsubsection{G036.914+0.489}

We re-observed G036.914+0.489 in order to resolve confusion about its
velocity.  L89 found two RRL velocities for his G036.914+0.489 target:
$-30.3$ and $+47.0$\,\kms.  There are in fact two distinct radio and
IR peaks associated with this region: G036.870+0.462\hide{LA082} and
G036.914+0.485\hide{LA082b}.  The velocities of both HRDS sources are
near $-30$\,\kms, implying a large distance from the Sun of
$\sim17$\,kpc (Anderson et al., 2011b, in preparation).  We do not
detect a second emission component from either source.

\subsubsection{Coincidence with Infrared Dark Clouds}

Some HRDS nebulae coincide with the locations of IR dark clouds
(IRDCs).  IRDCs are associated with the earliest phases of star
formation \citep{chambers09, rathborne10}.  If the nebulae and IRDCs
are truly spatially co-incident, then these regions are candidates for
the youngest sources in the HRDS.  The best examples of this
phenomenon are G009.875$-$0.749\hide{FA113},
G016.361$-$0.209\hide{LA1063}, G018.832$-$0.300\hide{LA409},
G027.366$-$0.164\hide{LA583},
G028.394+0.076\hide{LA598}, G029.770+0.219\hide{LA490},
G030.345+0.092\hide{LA008}, and G030.378+0.106\hide{LA009}, although
this is not a complete list.  These HRDS sources generally have a
small angular size and are frequently classified as ``compact.''  This
suggests a small physical size and an early evolutionary state. The
physical association between these HRDS nebulae and IRDCs, however, is
not always clear. For example, G028.394+0.076\hide{LA598} is entirely
surrounded by a very large IRDC.  We detect two velocities for this
source: 42.8 and 86.2\,\kms.  The 86.2\,\kms\ velocity is similar to
that of the IRDC, which is 75.8\,\kms\ \citep{simon06}.  The more than
10\,\kms\ difference is surprising, however, if the two are indeed
associated.  The nature of the relationship between this \hii\ region
and the IRDC remains unclear.

\subsubsection{Strange Line Profiles}

Two sources have double velocity component line profiles that are
different from all other HRDS targets.  G028.764+0.281\hide{LA485} has
hydrogen RRLs at 104.5 and 87.2\kms; G038.120$-$0.227\hide{LA100} has
its RRLs at 54.7 and 89.3\kms.  Both sources have line profiles with a
narrow component of line width $<13\,\kms$ and a broader component.
For G028.764+0.281\hide{LA485}, the narrow component is at
104.5\,\kms\ and has a FWHM line width of 12.5\,\kms.  For
G038.120$-$0.227\hide{LA100}, the narrow component has a velocity of
54.7\,\kms\ with a FWHM line width of 12.6\,\kms.  \citet{bronfman96}
measured a CS velocity of 83.5\,\kms\ for
G038.120$-$0.227\hide{LA100}, which implies that the narrow
54.7\,\kms\ line is from another source along the line of sight.  In
both cases it is not obvious what additional source along the line of
sight is producing such a narrow recombination line.  \cite{adler96}
argue, however, that the narrow RRL lines that they see in W\,3 are
caused by partially ionized hydrogen.


\subsubsection{Unusual Morphology}

One source, G349.579$-$0.680\hide{FA652}, has a morphology that is not
seen in any other HRDS target.  Its compact GLIMPSE
8.0\,\micron\ emission has the appearance of a bright-rimmed cloud,
whereas its MIPSGAL 24\,\micron\ emission is more extended and
diffuse.  G349.579$-$0.680 is near the center of a large (nearly
$1\,\arcdeg$ in diameter) 8.0\,\micron\ bubble located at \lb =
(349\,\fdg5, $-$0\,\fdg6).  There is no published RRL velocity for the
large bubble although it appears to be an \hii\ region based on
the mid-IR morphology.  The $-$19.4\,\kms\ velocity we measure for
G349.579$-$0.680\hide{FA652} may be that of the large bubble.

\subsubsection{Confusion with Other H{\scriptsize II} Regions}

In three cases an HRDS target may be part of a larger, more evolved
\hii\ region. For example, G347.536+0.084\hide{FA632} is part of the
ionization front of the large \hii\ region centered at \lb =
(347\,\fdg5, +0\,\fdg2).  \citet{caswell87} found RRL emission in two
locations along the ionization front at velocities of $-97$ and
$-96$\,\kms.  We measure a velocity of $-102.5$ for
G347.536+0.084\hide{FA632}.  One region, G049.507$-$0.520\hide{LA166},
has no GLIMPSE or MIPSGAL counterpart; it is a region of radio
continuum emission near W51 \lb = (49\,\fdg5,$-$0\,\fdg4).  Its
63.6\,\kms\ RRL velocity is near that of W51, for which
L89 measured a velocity of 58.2\,\kms.  Finally,
G006.014$-$0.364\hide{FA095} may be a second ionization front for
G005.899$-$0.427, for which L89 measured a velocity of
10.1\,\kms.  We measure a velocity for G006.014$-$0.364\hide{FA095} of
14.2\,\kms.

\subsection{Astrophysical Context\label{sec:correlation}}

%


Many of the HRDS sources appear in catalogs made from existing
Galactic surveys, but not as \hii\ regions.  The 448 nebulae detected
here in RRL emission for the first time, when correlated with previous
catalogs made at multiple wavelengths, can provide insight into the
properties of star forming regions.  Here we focus on the
astrophysical implications of HRDS associations with methanol masers
and dense molecular clouds traced by CS emission.

\subsubsection{Methanol Masers}

\hii\ regions are expected to be associated with methanol masers in
the earliest phases of high mass star formation \citep{ellingsen06}.
Although many of our HRDS nebulae have been observed in methanol maser
transitions, there are relatively few detections.  Using the
\citet{pestalozzi05} compilation of 14 methanol maser studies, we find
that only 10\% (46/448) of the HRDS sources have detected methanol
maser emission within $2\,\arcmin$ of their position.  Table
\ref{tab:sources} lists the velocity for the strongest maser
component.  This low correlation with methanol masers suggests that
either our sample of \hii\ regions is on average not extremely young
or that the maser emission is too weak to be detectable.  For
comparison, \citet{walsh98} found that 38\% (201/535) of the sources
in their UC \hii\ region sample have methanol maser emission.  This
suggests that the HRDS nebulae are not as young as their UC sample.
We defer a more detailed study of the evolutionary state of the HRDS
nebulae to a future publication.

\subsubsection{Dense Gas}


Tracers of dense molecular gas, such as CS, are hallmark
characteristics of young \hii\ regions \citep[e.g.,][]{churchwell90}.
The \citet{bronfman96} CS ($2\rightarrow1$) sample contains 110 of our
HRDS targets.  \citet{bronfman96} observed all IRAS point sources
satisfying the \citet{wc89b} IRAS color criteria for UC \hii\ regions.
Roughly half (57/110) of these HRDS sources were detected in CS.
Overall, the velocities are in good agreement; the mean difference
between the CS and RRL velocity is $4.2\,\pm\,5.9\,\kms$.  There are,
however, four sources with velocities that differ by more than
$10\,\kms$: 
G023.585+0.029, \hide{LA433} G028.304$-$0.390, \hide{LA478}
G349.437+1.058, \hide{FA002} and G000.382+0.017 \hide{FA070}.
Furthermore, there are two double-velocity RRL sources whose CS
velocity lies in between the RRL velocities: G352.521$-$0.144
\hide{FA017} and G049.998$-$0.125 \hide{LA276}.  Excluding these
sources, the average velocity difference is $3.0\,\pm\,2.3\kms$.
Thus, with some exceptions, the CS and RRL velocities are in very good
agreement.

The CS non-detections are equally interesting since they imply that
the \hii\ region is evolved and has displaced or dissociated its dense
molecular gas.  IRAS point sources satisfying the color criteria
employed by \citet{bronfman96} have generally been assumed to be UC
\hii\ regions; in many cases this assumption has been proven
valid \citep[e.g.][]{wc89a}.  That half of the HRDS sources targeted
were not detected in CS, however, implies instead that many of the
HRDS nebulae are distant and evolved.  The CS sensitivity limit may
have precluded the detection of faint CS emission from distant
nebulae.


\section{Properties of the HRDS Nebulae\label{sec:properties}}

We derive here the statistical properties of the HRDS sources.
Although the HRDS found 448 \hii regions, the number of physically
distinct nebulae represented by the 603 discrete hydrogen RRL
components is not well known.  Thermal radio sources can often be
resolved into several apparently physically distinct emission regions,
each having a somewhat different position and RRL velocity.
Furthermore, the most massive star-forming complexes, W43 for
example, are extended and fragmented into many sub-clumps of localized
star formation, which together can ionize a very large zone.  Many of
our multiple velocity component targets may be detecting such low
density ionized gas in addition to RRL emission from another,
physically distinct nebula (see \S\,\ref{sec:multiples}).  
Nonetheless, here we follow the convention established by L89 and
assume that each of the HRDS 603 RRLs is produced by a distinct
object.

We also wish to compare the properties of the newly discovered HDRS
nebulae with those of the previously known \hii\ region census.  There
is, however, no extant \hii\ region catalog that is suitable for this
purpose.  We have therefore compiled a catalog of previously known
\hii\ regions from existing RRL surveys \citep{reifenstein70, downes80,
caswell87, lockman89, lockman96, araya02, watson03, sewilo04}.  This
catalog improves upon previous efforts not only because it uses the
most recent observations but also because it accounts for duplicate
sources.  

Many \hii\ regions appear in multiple catalogs, so extreme care must
be taken when combining them into a single compilation.  We visually
inspect the radio continuum and MIR emission from all previous RRL
observations of HII regions to help identify sources that were
observed by multiple authors.  If observed positions are part of the
same contiguous radio continuum and MIR zone of emission and have RRL
velocities within 5\,\kms of one another, we deem them to be
observations of the same object.  For such matches we only keep the
source data from the most recent catalog, because the newest
recombination line data are almost invariably the most accurate.  We
have also removed known SNRs, LBVs, and PNe from the sample through
correlation with the SIMBAD database using the nominal HRDS target
positions and a 5\,\arcmin\ search radius.

We refer hereafter to this catalog of \hii\ regions previously known
to reside within the HRDS longitude and latitude range as the
``Known'' sample.  For each source, the Known sample catalog contains
the Galactic longitude and latitude, the RRL LSR velocity, and the
FWHM line width.
The Known sample data can be found online as an ASCII text file
\footnote{http://go.nrao.edu/hrds.  This catalog is as yet a work in progress. 
It still is likely to contain some residual contamination and duplicate entries.}; 
Anderson et al. (2011a, in preparation) give the full details.   
There are 456 unique \hii\ regions in the Known sample,
although there is likely some residual contamination.  With its 448
unique positions, the HRDS has thus doubled the sample of
\hii\ regions in the survey zone.

\subsection{Galactic Distribution\label{sec:galactic_structure}}

\begin{figure} \centering
\includegraphics[width=6 in]{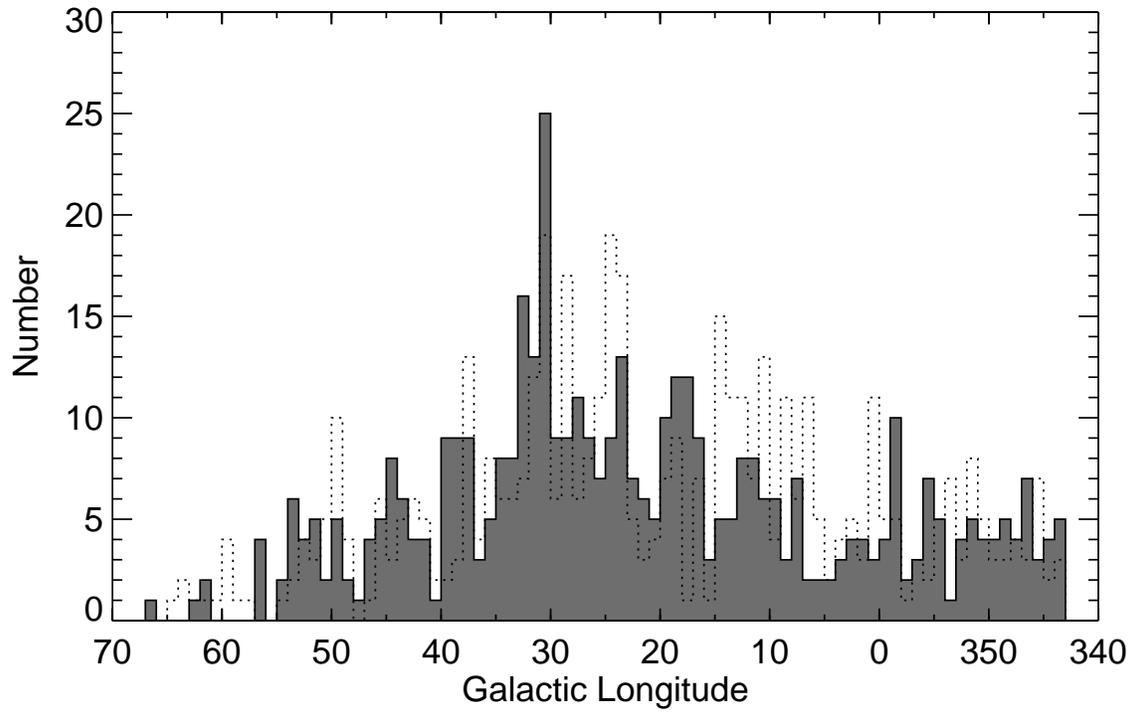}

\caption{Distribution of \hii\ regions in Galactic longitude.  
Histograms show the distribution for the HRDS (dark gray) and the Known sample
nebulae (dotted outline).  
}
\label{fig:hii_glong}
\end{figure}

\begin{figure} \centering
\includegraphics[width=5 in]{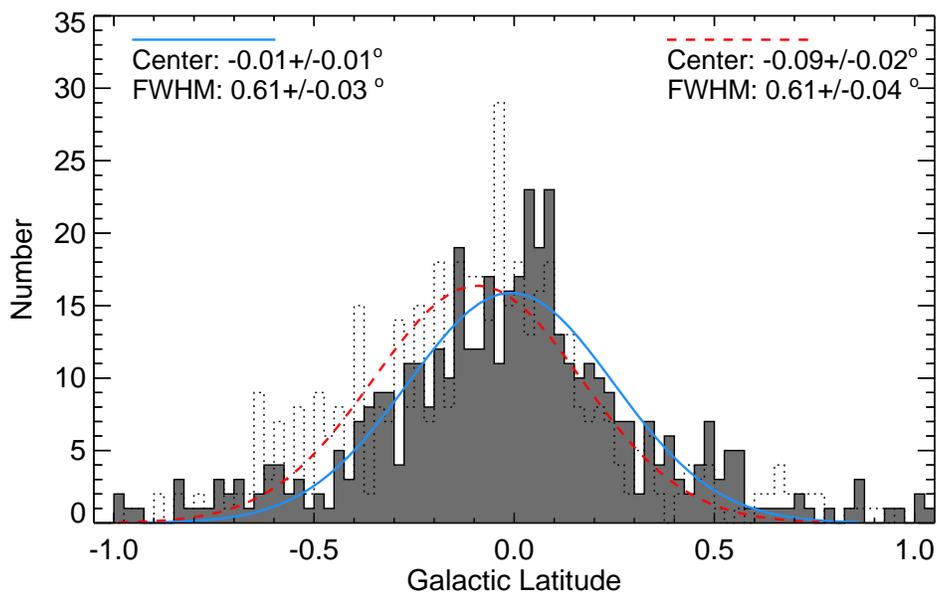}

\caption{Distribution of \hii\ regions in Galactic latitude.
Histograms show the distribution for the HRDS (dark gray) and the
known sample nebulae (dotted outline).  Here we include only nebulae  
that are located within 1\degree\ of the Galactic plane. Gaussian
fits to the two histograms are superimposed.  The two samples have
the same $\sim\,0\,\fdg6$ FWHM distribution in latitude.  The HRDS
source distribution is centered on the Galactic plane ($-0\,\fdg01$
is a displacement from \gb = $0\arcdeg$ of less than 1/2 the GBT's
HPBW), whereas the center of the known \hii\ region sample
distribution lies $\sim\,5\arcmin$ to the South.
}
\label{fig:hii_glat}
\end{figure}

The Galactic longitude distribution of the HRDS and Known samples,
shown in Figure~\ref{fig:hii_glong}, are very similar.  In
Figure~\ref{fig:hii_glong}, the HRDS sample is shown as dark gray and
the Known sample is shown as a dotted outline.  A Kolmogorov-Smirnov
(K-S) test shows that the distributions are not statistically
distinct.  The HRDS Galactic latitude distribution, shown in
Figure~\ref{fig:hii_glat}, peaks closer to the Galactic mid-plane
compared with the Known sample.  A Gaussian fit to the two
distributions shows that both the HRDS and the Known sample have essentially
the same FWHM scale-height of $0\,\fdg6$.  A K-S test of the two samples,
however, reveals significant statistical differences, entirely due to
the difference in peak latitude.  It is not clear why the two
distributions peak at different latitudes.  \citet{anderson_thesis}
found that the nebulae in our pilot study for the HRDS were on average
more distant than the Known sample \hii\ regions.  The HRDS may simply
be sampling a different area of the Galaxy.

\subsection{Galactic Structure\label{sec:velocity}}

The fundamental result of any RRL survey is the LSR velocity of a
nebula.  Thus the fundamental map of the observed data is its
\lv\ diagram.  The HRDS and the previously known \hii\ regions
together give a census of the Inner Galaxy, the zone within
$70\degree$ of the direction to the Galactic center, that shows
clear evidence for an ordered pattern of Galactic structure in
\lv\ space \citep{bania10}.  This was asserted by previous
researchers, but our new census, which has a factor of two more
objects, now unambiguously shows the kinematic signatures of the
spatial distribution of \hii\ regions, a concentration of nebulae at
the end of the Galactic bar, and nebulae located on the kinematic
locus of the 3 kpc Arm.
The HII region Galactocentric radial distribution shows very narrow
($\Delta\rgal\lsim1\,\kpc$) concentrations of star formation at
Galactic radii of 4.25 and 6.00\,\kpc.  This was known before but the
HRDS sharpens the contrast and now makes this structure very robust
\citep{bania10}.  More detailed Galactic structure studies that use
kinematic distances for HRDS nebulae determined by Anderson et
al. (2011b, in preparation) are underway (Bania et al. 2011, in
preparation).

The HRDS has 34 negative-velocity \hii\ regions in the first Galactic
quadrant at $\gl \ge 10\degree$.  This represents 7\% of all HRDS
nebulae in this zone.  First-quadrant sources with negative velocities
are unaffected by the kinematic distance ambiguity; assuming circular
rotation about the Galactic center, a negative RRL velocity implies
that a nebula is beyond the Solar orbit at a Galactocentric radius of
more than 8.5\,\kpc.  Before this survey, there were only seven known
\hii\ regions with negative-velocities in this region.
We have therefore increased the sample of negative-velocity
\hii\ regions thought to lie beyond the Solar orbit in this Galactic
zone by almost 500\%.  The negative-velocity sources are fainter on
average when compared with the HRDS sample as a whole.  Their average
continuum intensity is $\sim 250$\,mK, whereas the HRDS average is
$\sim 300$\,mK.  One negative-velocity source, G032.928+0.607,
\hide{LA039} has a continuum intensity of $\sim 700$\,mK, making it
one of the brightest HRDS nebulae.  Because of their extreme
distances, these sources are quite luminous.
They lie along the \lv\ locus of the Milky Way's ``Outer Arm''
\citep{bania10}.  These regions of massive star formation in the outer
Galaxy can provide important constraints on Galactic chemical
evolution.

The HRDS has nine sources in the Galactic center direction with RRL
velocities smaller than $-150$\,\kms.  This suggests that they lie in the
``Nuclear Disk,'' the region of high velocity gas in the inner $\sim
0.5$\,\kpc of the Milky Way.  These sources all lie between
$358\,\fdg5 < \gl < 359\,\fdg2$ and $\absb < 0\,\fdg1.$
\citet{lockman96} found two ``diffuse'' nebulae in this part of the
sky, which are the only previously known nebulae with similar
velocities.  We find only one source, G000.729$-$0.103, \hide{FA074} 
with a velocity, 105.3\,\kms, that could place it in the red-shifted
side of the nuclear disk.  Although these ten sources show that star
formation is ongoing in the nuclear disk,  \citet{caswell10} find no
methanol masers with nuclear disk velocities.

\subsection{Radio Recombination Line Widths \label{sec:linewidth}}

\begin{figure} \centering
\includegraphics[width=6 in]{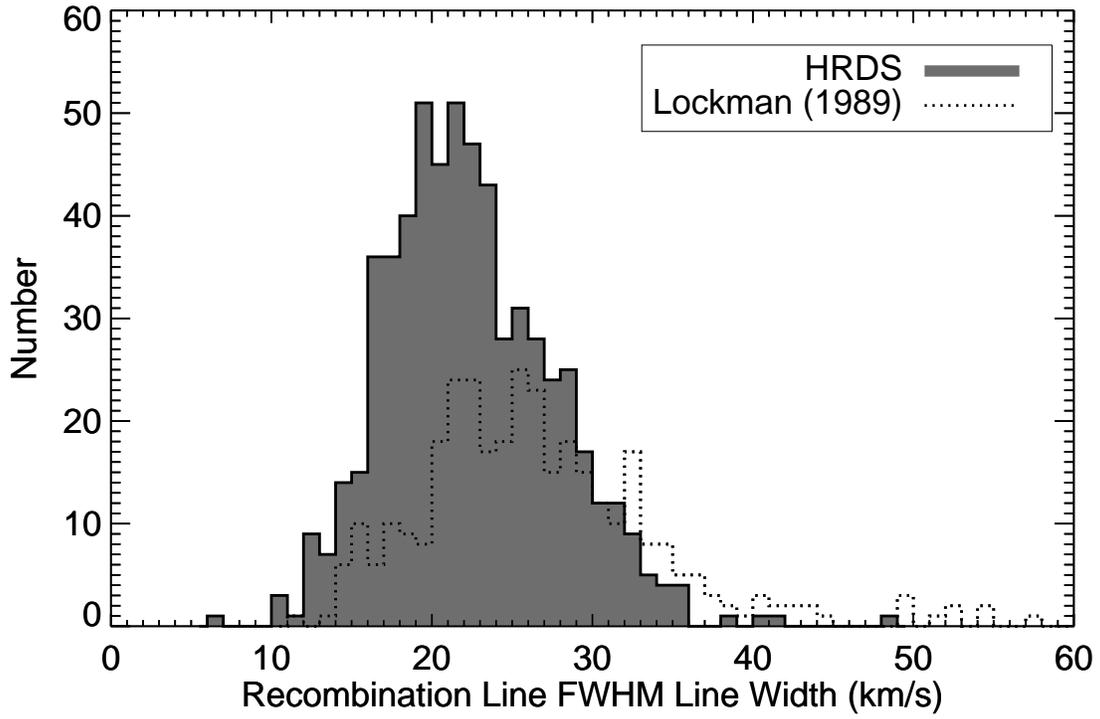}

\caption{Distribution of \hii\ region radio recombination line FWHM line widths.  
Histograms are shown for the HRDS (dark gray) and \citet{lockman89} (dotted outline)
\hii\ region samples.  On average the HRDS nebulae have narrower lines than the  
Lockman sources.  There are, moreover, many fewer HRDS nebulae with line widths 
greater than 35\,\kms.}
\label{fig:fwhm}
\end{figure}

The $\sim25\kms$ line widths of \hii\ region RRLs are set by a
combination of thermal, turbulent, and ordered motions (e.g., flows).
This line width greatly exceeds the thermal width expected from the
$\sim\expo{4}\K$ temperature that is typical of \hii\ region
plasmas.  \hii\ region RRLs are thus significantly broadened by
turbulent and ordered motions.  In Figure~\ref{fig:fwhm}, we show the
RRL full width at half maximum (FWHM) line width distribution for the
HRDS sample (dark gray) and the subset of the L89 sample from the HRDS
Galactic longitude and latitude range (dotted outline).  We compare
the HRDS with a single sample, that of L89, rather than the Known sample
in order to minimize any possible systematic effects stemming from
different observing and data analysis techniques.

The typical FWHM line widths for the HRDS nebulae are narrower than
those of the L89 \hii\ regions.  The distribution for the HRDS sample
peaks at $\sim20$\,\kms, whereas the L89 sample is noticeably broader
and shows peaks at 23 and 26\,\kms.  The formal means and standard
deviations are $22.3\pm5.3$ for the HRDS compared with $26.4\pm8.1$
for L89.  A KS test of the HRDS line widths (603 sources) and the L89
line widths (462 sources) shows that the two distributions are
statistically distinct to a high degree of certainty.
Some of the difference between the two
distributions may be due to the different spectral resolutions, which
are 1.86 and 4\,\kms, respectively, for the HRDS and L89 surveys.  The
L89 line widths could also be broader because the 140 Foot telescope's
bigger beam (195\,\arcsec\ compared with the GBT's 82\,\arcsec~)
sampled a much larger volume of gas. 

The HRDS has 35 nebulae with extremely narrow, $<15\,\kms$, line
widths.  These line widths can be used to estimate the maximum electron
temperature of the \hii\ region plasma.  Since turbulence and ordered
motions only broaden the RRLs, interpreting the line width as being
purely thermal sets a robust upper limit on the nebular electron
temperature.
Extremely narrow line widths, $\lesssim15 \kms$, imply ``cool''
nebulae with electron temperatures $\lsim5,000\,$K.  Such nebulae were
discovered by \citet{shaver79b}.  They were first detected in
appreciable numbers by L89, who speculated that cool nebulae may be
common and that few were known because of the sensitivity limits of
the existing RRL surveys.  The HRDS, however, shows that cool nebulae
are indeed rare in the Galaxy: only $6\%$ of HRDS sources have line
widths smaller than $15$\,\kms.  This fraction is, however, three
times higher than that found for the L89 sample.


We detect only three nebulae with large, $\gtrsim 40$\,\kms, line
widths.  Ignoring any contribution from ordered motions and assuming
that the thermal and turbulent components add in quadrature, lines
broader than $40\,\kms$ imply turbulent motions of order 35\,\kms\ for
a typical $10,000\K\,$ \hii\ region plasma.  Some broad line width
sources, however, may in fact be PNe (see \S\ref{sec:indiv_sources}).

\begin{figure} \centering
\includegraphics[width=6 in]{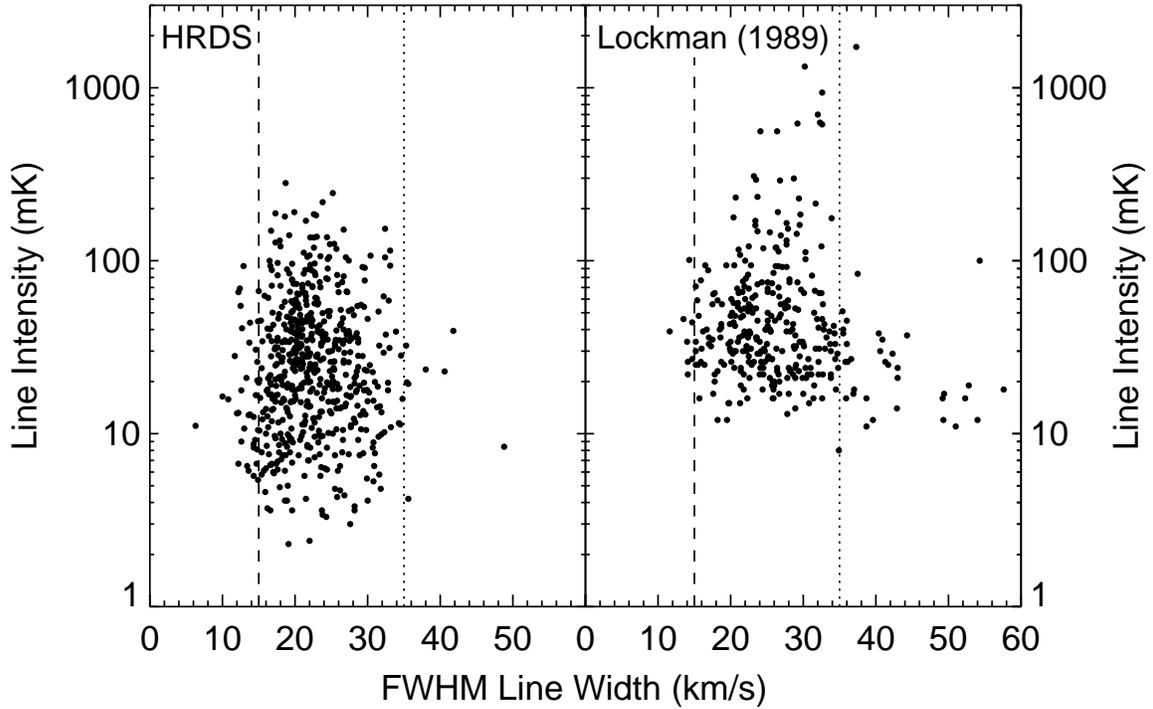}

\caption{\hii\ region RRL line intensity as a function of FWHM line width.  
Shown are the HRDS (left panel) and L89 samples (right panel).  
Vertical dotted lines mark line widths of 35\,\kms.
Low intensity HRDS sources share the same distribution of line widths as
the entire sample.  The L89 sources, however, show that systematically 
broader lines are found for low intensity sources.  Vertical dashed lines
mark line widths of 15\,\kms.  Nebulae with line widths this narrow are 
``cool'';  they cannot have electron temperatures greater than 5,000\,\K.
}
\label{fig:height_vs_fwhm}
\end{figure}

As the spectral signal-to-noise decreases, the line width derived from
Gaussian fitting becomes increasingly uncertain.  Furthermore, with
low signal-to-noise, double-velocity components may be not be
well-separated and an erroneously large line width might be derived.
In Figure~\ref{fig:height_vs_fwhm}, we plot the line width as a
function of the line intensity for the HRDS (left panel) and the L89
samples (right panel).  Figure~\ref{fig:height_vs_fwhm} shows that the
line width distribution for low intensity HRDS sources is similar to
that of the entire distribution.  This is not the case for the L89
sample --- many broad line sources are of low intensity.
We conclude that at centimeter-wavelengths Galactic \hii\ regions have
line widths $\le 35$\,\kms.  The HRDS has only eight lines (1\%)
broader than $35$\,\kms, and only four lines broader than $36$\,\kms.
Line widths greater than 35\,\kms\ should be regarded with suspicion
because they may be the result of a low signal-to-noise detection,
represent blended velocity components, or a sign that the source is a
PN and not an \hii\ region.

\subsection{Continuum Angular Size \label{sec:size}}

The distribution of the continuum angular sizes for HRDS nebulae is
shown in Figure~\ref{fig:continuum_size} which plots the geometric
mean of Gaussian fits to the RA and Dec continuum scans.  Altogether,
we have a useable continuum measurement in either RA or Dec. for 441
HRDS nebulae.  Here we use the fitted FWHM angular sizes uncorrected
for the GBT beam.  Typical errors for the fits are 5 to 10\,$\arcsec$.
Assuming Gaussian source and beam shapes, the observed and true
angular sizes are related by the expression:
$\theta^2 = \theta_s^2 + \theta_b^2\,$,
where $\theta$ is the observed size (shown in
Figure~\ref{fig:continuum_size}), $\theta_s$ is the true source size,
and $\theta_b$ is the beam size. The vertical dashed line in
Figure~\ref{fig:continuum_size} flags the 116\,\arcsec\ size that would
be measured by the GBT for a source that just fills its X-band beam.
After correcting for the beam size, the majority of the HRDS nebulae,
235 of 441 (53\%), are unresolved by the $82\,\arcsec$ GBT beam. 
Eleven sources ($\sim5\%$) have angular sizes larger than 
$250\,\arcsec$; these are not shown in Figure ~\ref{fig:continuum_size}.

Although small in angular size, many HRDS nebulae lie at extreme
distances from the Sun \citep[][Anderson et al., 2011b, in
preparation]{anderson_thesis}.  Thus many of these angularly small
nebulae are actually large in physical size.  If the HRDS nebulae have
on average large physical sizes, this would imply an evolved state and
would explain why the HRDS targets are in general not correlated with
methanol masers and CS emission (see \ref{sec:correlation}).

\subsection{Planetary Nebula Contamination\label{sec:PNe}}

There is
likely some contamination from PNe in the HRDS sample because our
selection criteria locate {\it thermally} emitting sources and not
just \hii\ regions.  We have, however, identified only 17 HRDS sources
that might be PNe (see \S~\ref{sec:PNeNote}).  Thus fewer than 4\%
(17/448) of the HRDS sources are potential PNe candidates, which
confirms \citet{bania10} who argued that the level of contamination in
the HRDS by PNe must be quite small.  This conclusion was based on
considerations of Galactic structure, scale height, RRL line widths,
and the RRL line-to-continuum ratios (i.e. nebulae electron
temperatures).
In particular, the clear evidence for Galactic structure seen in the
\lv and Galactic radial distributions of HRDS nebulae implies that the
level of PNe contamination in the HRDS sample must be minimal.
Because PNe are an old stellar population their Galactic orbits are
well-mixed.  PNe show, therefore, no structure in their Galactocentric
radial distribution and their Galactic \lv distribution is a scatter
plot constrained only by velocities permitted by Galactic rotation.
Any PNe contamination of the HRDS sample must therefore be very small,
otherwise these interlopers would suppress the unambiguous signal of
Galactic structure seen in the HRDS Galactocentric radial and \lv
distributions.

\subsection{Survey Completeness\label{sec:completeness}}


By assuming a flux distribution, one can assess the completeness of a
flux-limited survey.  There are good reasons to expect a power law for
the HRDS fluxes. For example, power law flux distributions have been
found for the MSX, GLIMPSE, and IRAS point source catalogs.  A power
law distribution of fluxes reflects an underlying power law for source
luminosities together with a relatively structureless pattern of
source locations across the Galaxy.  For \hii\ regions, it is
well-known that a power law is a good approximation to the luminosity
function \citep[e.g.,][]{smith89, mckee97} and that they are
relatively smoothly located across the Galaxy.  While there are strong
peaks in the \hii\ region {\it Galactocentric} distance distribution
\citep[cf.][]{anderson09a, bania10}, their {\it heliocentric} pattern
shows no structure.

We attempt to assess the completeness of the HRDS by evaluating the
integrated continuum flux distribution calculated using Equation 
\ref{eq:integrated_flux} and assuming a power-law distribution of
fluxes.

Because the radio continuum data suffer from confusion and possible
zero point uncertainties, we also use the RRL data to estimate the
continuum flux of the HRDS nebulae.  To do this we assume an X-band
line-to-continuum ratio of 0.09.  This ratio is the average value of
the \citet{quireza06a} sample of Galactic \hii\ regions, derived using
their 28 highest quality measurements (their quality factor values of
``A'' or ``B'' for both the line and the continuum data).  We then use
Equation~\ref{eq:integrated_flux} and the fitted RA and Dec continuum
scan widths to get the integrated continuum fluxes.  In both cases we
are assuming that the sources are Gaussian in both RA and Dec and
well-characterized by the cross scans.  We are also implicitly
assuming that the HRDS nebulae do not suffer greatly from optical
depth effects (see \S\ref{sec:optical_depth}).

\begin{figure} \centering
\includegraphics[scale=0.9]{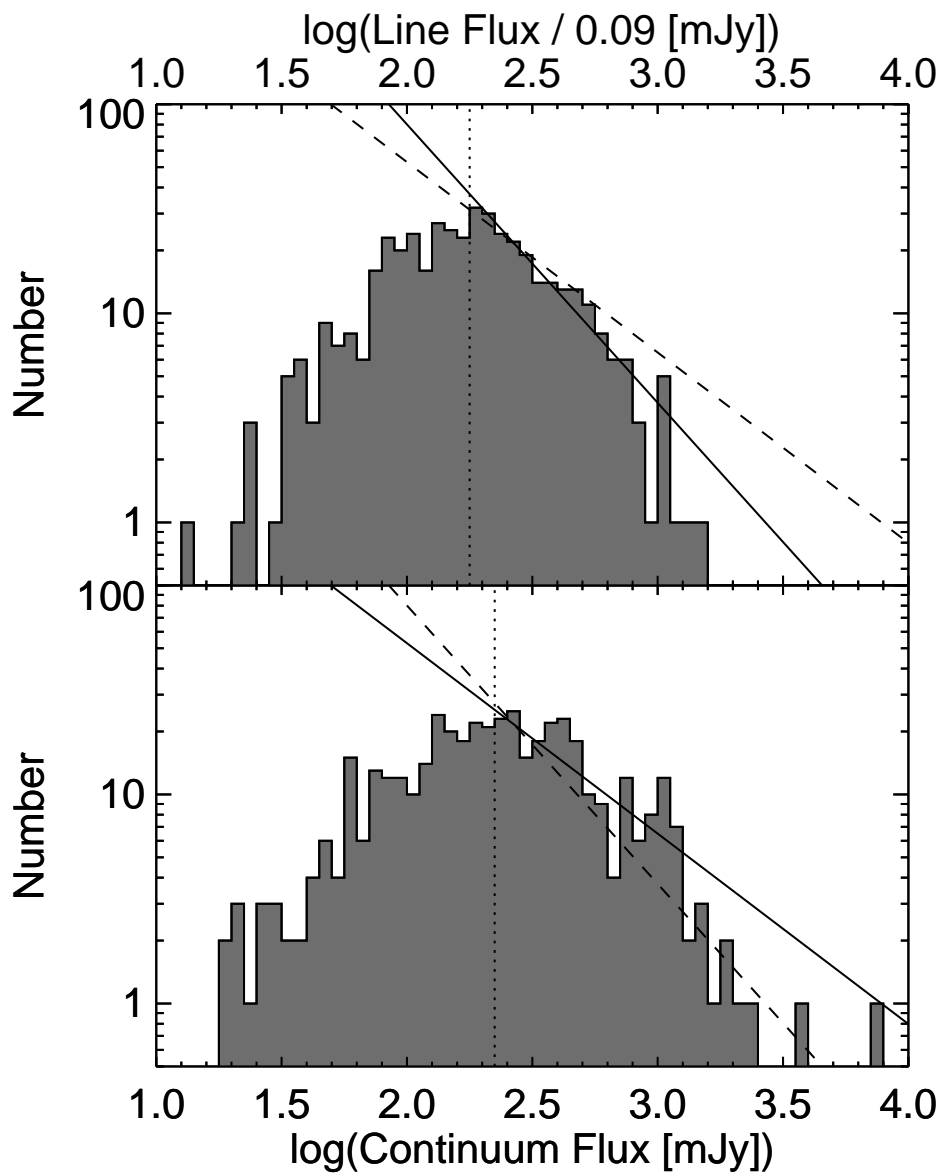}

\caption{Integrated source flux distributions for HRDS nebulae.  
The bottom panel shows the observed integrated continuum flux distribution. 
The top panel shows the integrated continuum flux distribution derived
from the observed RRL line intensities and the continuum angular
sizes, assuming a line-to-continuum ratio of 0.09.  Solid lines are
power law fits to each histogram.  Dotted lines flag the point where
the observed distribution deviates from the power law.  Dashed lines
show the power law fitted to the data in the other histogram. 
}
\label{fig:continuum_flux}
\end{figure}

Histograms of the integrated source flux distribution for HRDS nebulae
are plotted in Figure~\ref{fig:continuum_flux}, where the bottom panel
shows the fluxes derived using the continuum data and the top panel
shows continuum fluxes estimated from the RRL data.  Solid lines are
power-law fits to a panel's data for the range 200 to 1,000\,mJy.  The
dashed line shows the fit to the other panel's data.  We estimate the
completeness limit using the histogram bin where the fit deviates
significantly from the fitted power law; these limits are shown as
vertical dotted lines.
The fit to the continuum flux distribution has a slope of $-0.8$ and
shows a completeness limit of $\sim 280$\,mJy.  This fit is, however,
rather poor (reduced $\chi^2$=15).  The fit quality of the RRL
estimated continuum flux is much better (reduced $\chi^2$=3).  It is
best-fit with a slope of $-1.0$ and shows a completeness limit of
$\sim 180$\,mJy.  The HRDS completeness limit of 120\,mJy cited by
\citet{bania10} was based on an early analysis of sources from $\gl =
30\arcdeg$ to $50\arcdeg$.  It appears that the completeness limit is
lower in this region of the Galaxy, possibly because these candidate
targets were identified using the more-sensitive VGPS data.

There is a large difference between the two distributions shown in
Figure~\ref{fig:continuum_flux}.  Some of the discrepancy is caused by
assuming a single line-to-continuum ratio for the entire sample.  In
reality the line-to-continuum ratio varies between \hii\ regions.
Perhaps a larger issue, however, is that many of our targets are in
locations with significant continuum emission structure.  It is often
difficult to separate the target's emission from that of other sources
and/or noise due to sky fluctuations.  In many cases it is also
difficult to establish a reliable zero point for the continuum
intensity.  Because of these concerns, even though we must assume a
line-to-continuum ratio, using the line intensity as a proxy may in
fact be a better estimate of the actual continuum flux of faint
sources.

The completeness limit derived here is over twice that expected from
our HRDS target list.  This discrepancy almost certainly stems from
the uncertainty in the extrapolation of 20\,cm radio continuum fluxes
to predicted 3\,cm (X-band) GBT fluxes.  We found empirically that the
flux density actually measured at X-band can differ by a factor of two
from the extrapolated flux density.


\subsubsection{Spectral Type Completeness\label{sec:spectral_type}}

At X-Band, RRL emission propagates on average through an optically
thin medium and thus we can detect \hii\ regions on the far side of
the Galaxy.  Here we estimate the distance to which we can detect
\hii\ regions ionized by stars of various spectral types given the
derived HRDS completeness limit of 180\,mJy.  To do this we first must
calculate the luminosity dependence of \hii\ regions for a range of
ionizing spectral types.

The spectral type of the exciting star of an \hii\ region, assuming
ionization by a single star, can be estimated from its radio flux
\citep{rubin68}:
\begin{equation} 
N_{\rm ly} \approx 4.76 \times 10^{48} \left(\frac{S_\nu}{\rm Jy}\right)
\left(\frac{T_{\rm e}}{\rm K}\right)^{-0.45} 
{[\rm\,s^{-1}]}\,,
\label{eq:n_ly_rubin}
\end{equation}
where $N_{\rm ly}$ is the ionization rate, the number of Lyman Continuum ionizing photons 
emitted per second, $S_\nu$ is the nebular radio flux density, and $T_e$ is the electron 
temperature. 
This equation is almost unaffected by source geometry and variations
in electron density within the source, but does assume that the continuum emission
is optically thin.

\begin{figure} \centering
\includegraphics[scale=0.9]{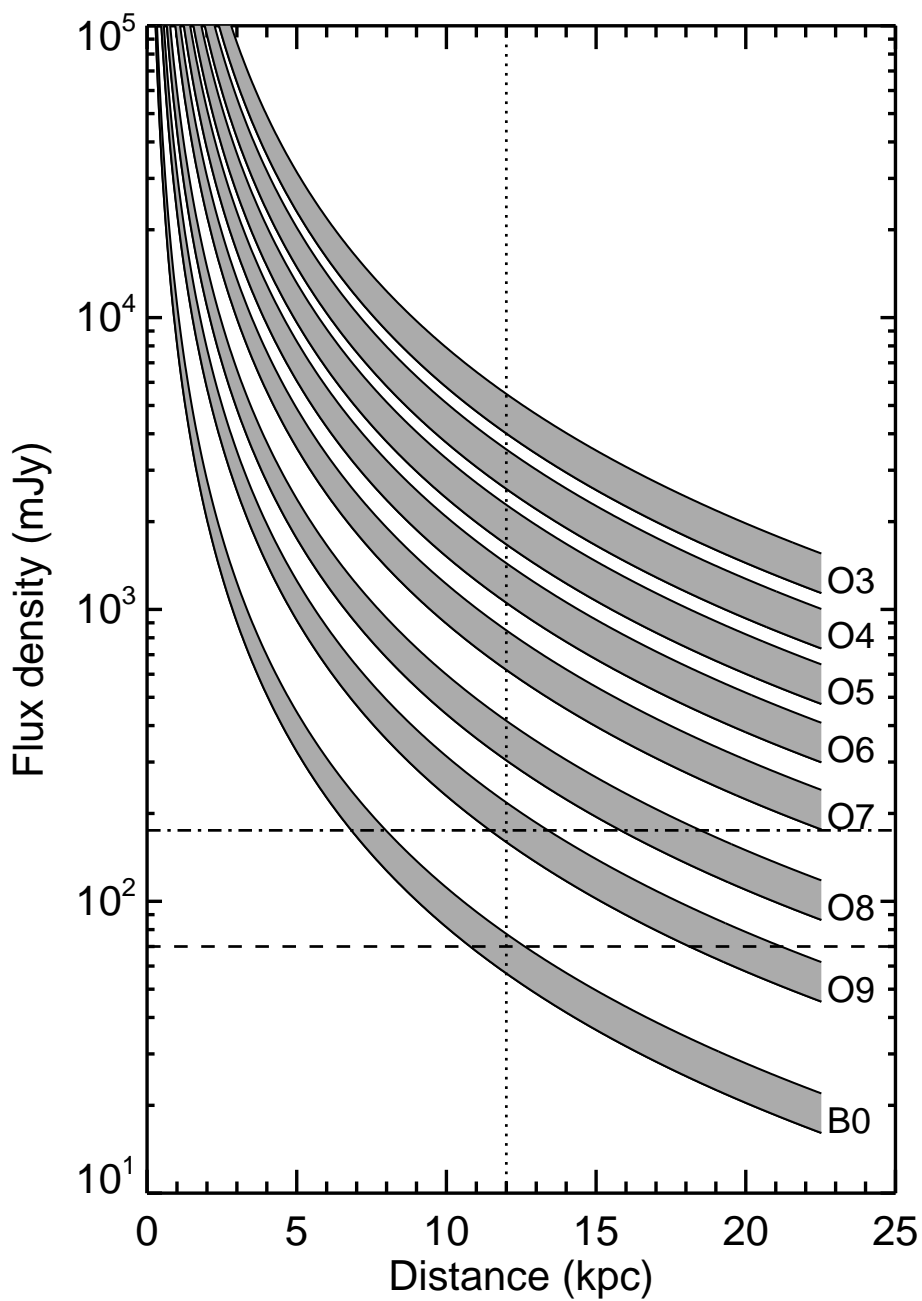}

\caption{Flux of \hii\ regions ionized by single OB stars as a function
of distance and spectral type.  The gray curves show the estimated
flux for \hii\ regions with electron temperatures in the range 5,000
to 10,000\,K.  The dashed line shows the $\gtrsim\,70\,\mjy$
9\,GHz extrapolated flux limit of the HRDS target list, whereas the dash-dot 
line shows the $\gtrsim\,180\,\mjy$ flux limit derived from the 
survey itself (see \S\,\ref{sec:completeness}).  This flux limit implies 
that the HRDS is complete for \hii\ regions ionized by single O-stars within 
12\kpc of the Sun (dotted line).}  
\label{fig:flux_v_distance}
\end{figure}

Using the Lyman continuum photon emission rates for stars of spectral
type O3 through B0.5 calculated by \citet{sternberg03}, we convert the
Equation \ref{eq:n_ly_rubin} $N_{\rm ly}$ values into ionizing star
spectral types.  We plot in Figure~\ref{fig:flux_v_distance} the
expected flux density of \hii\ regions ionized by single stars of
spectral type O3 to B0 as a function of the distance to the
\hii\ region.  The gray curves in Figure~\ref{fig:flux_v_distance}
show the estimated flux for the range of electron temperatures found
in \hii\ regions, $\sim5,000$ to $\sim10,000$\,K.  The dashed line
shows the 70\mjy extrapolated 9\,GHz flux limit of the HRDS target list.  The
dash-dot line shows the actual HRDS 180\mjy flux limit derived in \S
\ref{sec:completeness}.  The HRDS is complete for all \hii\ regions
surrounding single O-stars within 12\kpc of the Sun (dotted line).
Being flux-limited, however, the HRDS is complete to different
distances for each spectral type.  The HRDS can detect \hii\ regions
ionized by single O7 stars across the entire Galactic disk. Since the
ionizing flux decreases rapidly for later spectral types, however, we
can only detect \hii\ regions ionized by B1 stars at a distance of
only a few \kpc.

\subsubsection{Optical Depth Effects and Confusion\label{sec:optical_depth}}

Young \hii\ regions may be optically thick at 20\,\cm, an effect which
could affect our completeness.  We use the 20\,\cm VGPS and NVSS data
to estimate the X-band flux.  If a nebula is optically thick at
20\,\cm then its extrapolated X-band flux would be underestimated.
The source might then fall below our flux threshold and so not make it
onto the HRDS target list.
The flux density of an \hii\ region rises as $\nu^2$ until it peaks
at a frequency, $\nu_0$, called the ``turnover frequency.'' At
frequencies less than $\nu_0$ the nebula is optically thick whereas at
frequencies greater than $\nu_0$ it is optically thin.  In the
optically thin regime, the flux density has a very weak dependence on
frequency, $S\propto\nu^{-0.1}$.  The turnover frequency is:
\begin{equation}
\nu_0 = 0.3045\, \left(\frac{T_{\rm e}}{\rm K}\right)^{-0.643}
\left(\frac{EM}{\rm cm^{-6}\,pc}\right)^{0.476}\,[{\rm GHz}]\,,
\label{eq:turnover}
\end{equation}
where EM is the emission measure and $T_e$ is the electron
temperature.  For compact \hii\ regions with $T_e\approx 10^4$\,K and 
${\rm EM} \approx 10^6\,{\rm cm^{-6}\,pc}$, the turnover frequency is
near 1\,GHz.  Some UC and hyper-compact \hii\ regions may have emission
measures in excess of $10^9$ \citep{keto08}, which leads to turnover
frequencies near 15\,GHz.  Thus, optical depth effects become
important for young UC \hii\ regions with high emission measures and
for observations at low frequencies.  Our survey therefore is less
complete for the youngest UC and hyper-compact \hii\ regions that are
optically thick at 1.4\,GHz because we have not targeted these nebulae
in our observations.

While we make no attempt to prove empirically that our target location
method finds all possible \hii\ region candidates, it does not seem to
be severely affected by non-detections in the radio and IR surveys or
by source confusion. In the HRDS survey zone we find over 1,000
\hii\ region candidate targets that have spatially coincident MIR and
radio continuum emission with a similar morphology and angular extent.
About 40\% of these objects have 20\,cm flux densities {\it below} the
85\,mJy HRDS 20\,cm flux density cutoff.  Our method thus easily
finds a large number of sources fainter than the flux threshold we
used for the HRDS.  We conclude that the apparent brightness of
sources in the radio and IR data does not severely impact our
completeness.

Confusion may also limit the completeness of this survey.  Because of
the large number of sources found near W43 at $(l, b) =
(30\,\fdg78, $-$0\,\fdg03)$, which is the most complicated emission
region within the survey zone, we do not believe confusion
significantly affects our ability to visually identify \hii\ region
candidates.  In large part this is because that radio emission is
relatively unconfused compared to the IR emission.  The most
significant problem affecting our completeness is likely the errors in
the X-band fluxes extrapolated from the 20\cm radio continuum data
because this has a strong impact on which sources were observed.

\section{Summary\label{sec:summary}}

The Green Bank Telescope \hii\ region discovery survey at 9\,GHz has
doubled the number of known \hii\ regions in the Galactic zone
$343\arcdeg \le \gl \le 67\arcdeg$ with $\absb \le 1\arcdeg$.
The HRDS targets were selected based on spatially coincident
24\,\micron\ and 20\,cm emission of a similar angular extent.  Our
RRL detection rate of 95\% proves that this criterion is extremely
effective in identifying \hii\ regions.  The derived survey
completeness limit, 180\,mJy at X-band, is sufficient to detect all
optically thin \hii\ regions ionized by single O-stars at a distance
of 12\kpc from the Sun.  These recently discovered nebulae share the
same distribution on the sky as does the previously known census of
Galactic \hii\ regions.  On average, however, the new nebulae have
fainter continuum fluxes, smaller continuum angular sizes, fainter RRL
intensities, and smaller RRL line widths.

Nearly 30\% of the HRDS \hii\ regions (129 nebulae) have multiple
hydrogen RRLs in their spectra at different velocities.  Moreover,
over 30\% of these multiple velocity component sources are located
with 2\arcdeg of the large, massive star forming region W43.  Low
density ionized gas produced by FUV photons leaking into the extended
PDR surrounding W43 appears to be responsible for nearly all the
multiple velocity components detected in this region of sky.  We
speculate that the majority of the HRDS multiple velocity component
RRL spectra stem from similar low density PDR plasmas distributed over
a large region of star formation.  These zones are so large in angle
that they overlap the lines of sight to our HRDS targets.

The HRDS has 207 \hii\ regions with a bubble morphology at
8.0\,\micron\ in the GLIMPSE survey, 65 of which have been cataloged
previously.  By comparing the positions of bubbles detected in GLIMPSE
with the Known sample of \hii\ regions, we find that about half of all
Galactic \hii\ regions have a bubble morphology in the infrared.  We
speculate that nearly all GLIMPSE bubbles are caused by \hii\ regions
\citep{bania10}. The bubble morphology is probably not associated with
any particular evolutionary stage, but is rather an indication of the
homogeneity of the ambient environments of these \hii\ regions.  The
HRDS statistics on the distribution of different morphologies and the
line widths of bubble sources strongly suggest that these objects are
limb brightened projections on the sky of optically thin three
dimensional structures, rather than two-dimensional rings produced by
bipolar flows.

The HRDS has 34 \hii\ regions with negative LSR velocities.  These
first Galactic quadrant nebulae lie beyond the Solar orbit at \dsun
\gsim 12\kpc in the outer Galactic disk, \rgal \gsim 9\kpc, placing
them in the high-z, warped Outer Arm.  Previously there was only a
single \hii region known in this zone.  This new sample of outer
Galactic disk \hii\ regions paves the way for future studies.  Because
metals are the main coolants in the photo-ionized gas, the \hii region
heavy element abundance can be derived by proxy using the nebular
electron temperature, \te.  The electron temperature of \hii\ regions
is directly related to the amount of processing of the gas and
dust from which it formed, and thus to Galactic chemical evolution
(GCE).  There are relatively few \hii regions with accurately derived
\te values, especially for the \rgal\,$\gsim$\,10 \kpc region which is
critical for constraining models of GCE \citep{fu09}.  Determining electron
temperatures for these Outer Arm and other HRDS nebulae will be of
great importance for future GCE studies.

The HRDS also has ten \hii\ regions whose position and velocity place
them in the nuclear disk.  Previously there were only two
\hii\ regions known to lie in this part of the Galaxy.  These HRDS
nebulae show conclusively that active star formation is occurring in
the innermost part of the Milky Way.

The distribution of HRDS nebular line widths peaks at $\sim 20\kms$,
which is smaller than the $\sim 25\kms$ line widths found for the
previously known \hii\ region sample.  The combined census
distribution shows that Galactic \hii\ regions at 9\,GHz have line
widths smaller than 35\kms.  Line widths greater than this may be the
result of multiple components along the line of sight, a low
signal-to-noise detection, or misidentification of a planetary nebula
as an \hii\ region.  The HRDS has 35 nebulae with extremely narrow,
$<15\kms$, line widths.  These narrow line widths imply ``cool''
nebulae with electron temperatures $\lsim 5,000\K$.  Such nebulae were
first detected in appreciable numbers by \citet{lockman89}, who
speculated that cool nebulae may be common and few were known because
of survey sensitivity limits.  The HRDS shows, however, that cool
nebulae are indeed rare in the Galaxy: only 6\% of HRDS \hii\ regions
have line widths smaller than 15\kms.

Finally, the majority of the HRDS \hii\ regions are unresolved with
the $82\arcsec$ GBT beam at X-band.  Despite their small angular size,
however, it would be wrong to assume that these nebulae are in an
early stage of their evolution.  The HRDS sample shows a poor
correlation with dense gas and methanol masers, which are usually
associated with young \hii\ regions.  Instead, the HRDS nebulae
probably span a wide range of evolutionary stages.  Their small
angular sizes result from their being located on average at large
distances from the Sun.

\appendix
\section{The HRDS Web Site}
\begin{figure} \centering
\includegraphics[scale=0.65]{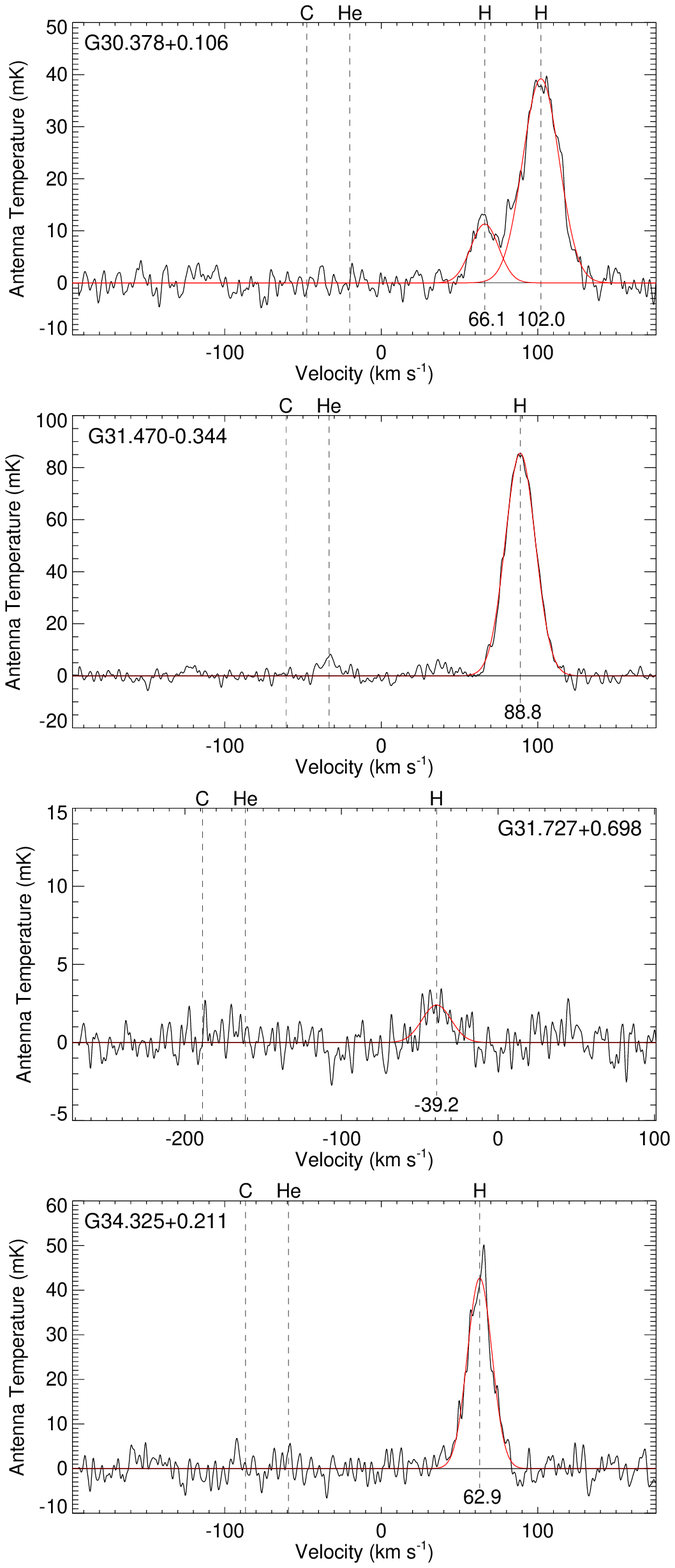}

\caption{Composite \hnaa\ RRL discovery spectra for the four Figure
\ref{fig:survey_examples} \hii\ regions.  Shown is the average of the
\hal{87} through \hal{93} RRLs, re-sampled to the resolution of the
\hal{87} spectrum and smoothed to 1.86\,\kms\ resolution.  A
Gaussian fit to each hydrogen RRL component is superimposed.  A
vertical dashed line flags the nebula's hydrogen RRL LSR velocity,
which is listed at the bottom of the flag.  The expected locations of
the helium and carbon RRLs are also flagged.  The G031.470$-$0.344
spectrum clearly shows a helium line.  For G030.378+0.106, hydrogen RRL
emission is seen at two velocities.  The 102.6\kms component is
probably associated with the W43 star forming complex (see
\S\ref{sec:multiples}).
}
\label{fig:example_lines}
\end{figure}

\begin{figure} \centering
\includegraphics[width=6 in]{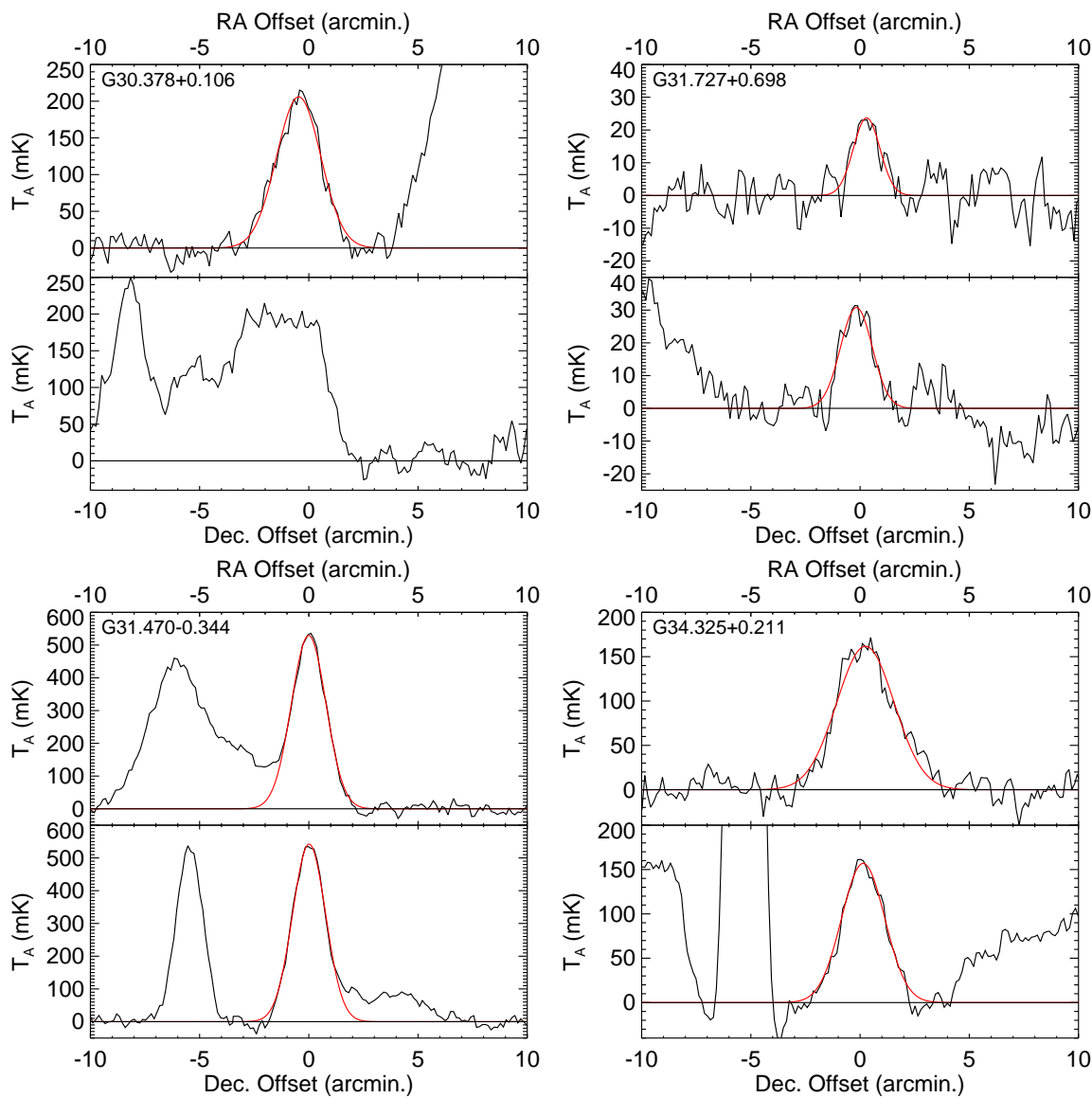}

\caption{Radio continuum measurements of the four Figure
\ref{fig:survey_examples} \hii\ regions.  Shown are the RA (top
panels) and Dec (bottom panels) position scans centered at the
nominal source position.  Gaussian fits to the continuum 
emission that we associate with the nebula are superimposed.  
These typical data show that the continuum emission in the inner 
Galactic plane is often extremely complex.  It is not always 
possible to distinguish between real continuum emission structure
and sky noise.}

\label{fig:example_continuum}
\end{figure}

\begin{figure} \centering

\includegraphics[scale=0.50]{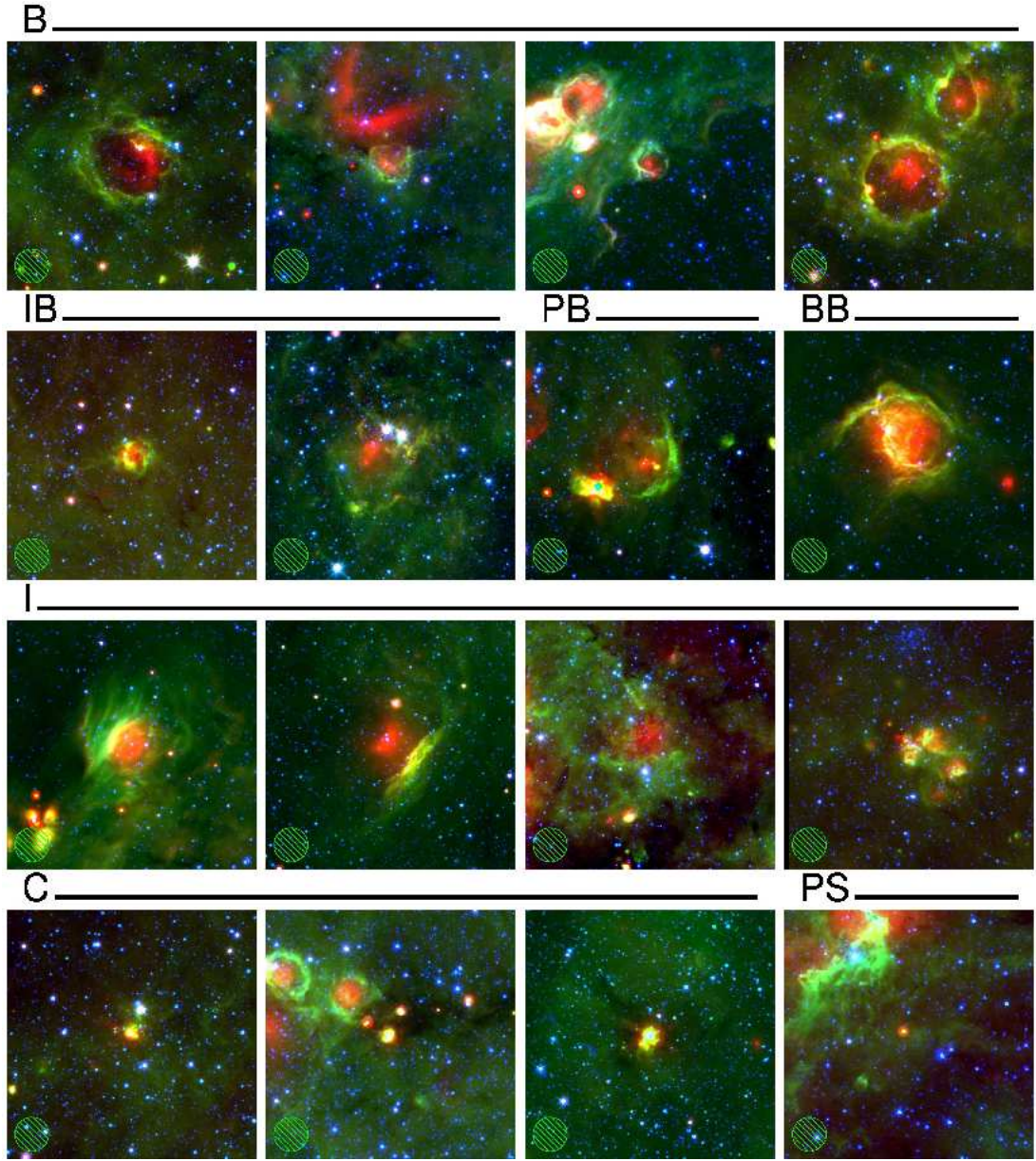}
\caption{Examples of HRDS target morphological classifications (see Table\,\ref{tab:sources}).  
Each panel shows a three-color image constructed from Spitzer infrared data: 
MIPSGAL 24\,\micron\ (red), GLIMPSE 8.0\,\micron\ (green) and GLIMPSE 3.6\,\micron\ 
(blue).  The HRDS morphological classification is based on
the target's appearance in the  8.0\,\micron\ image.  
Each image is $5\arcmin$ square and the $82\arcsec$ GBT beam is shown
in the lower left corner.  It is evident that the 8.0\,\micron\ emission always extends
beyond that of the 24\,\micron\ emission.}
\label{fig:threecolor}
\end{figure}

\begin{figure} \centering
\includegraphics[width=6 in]{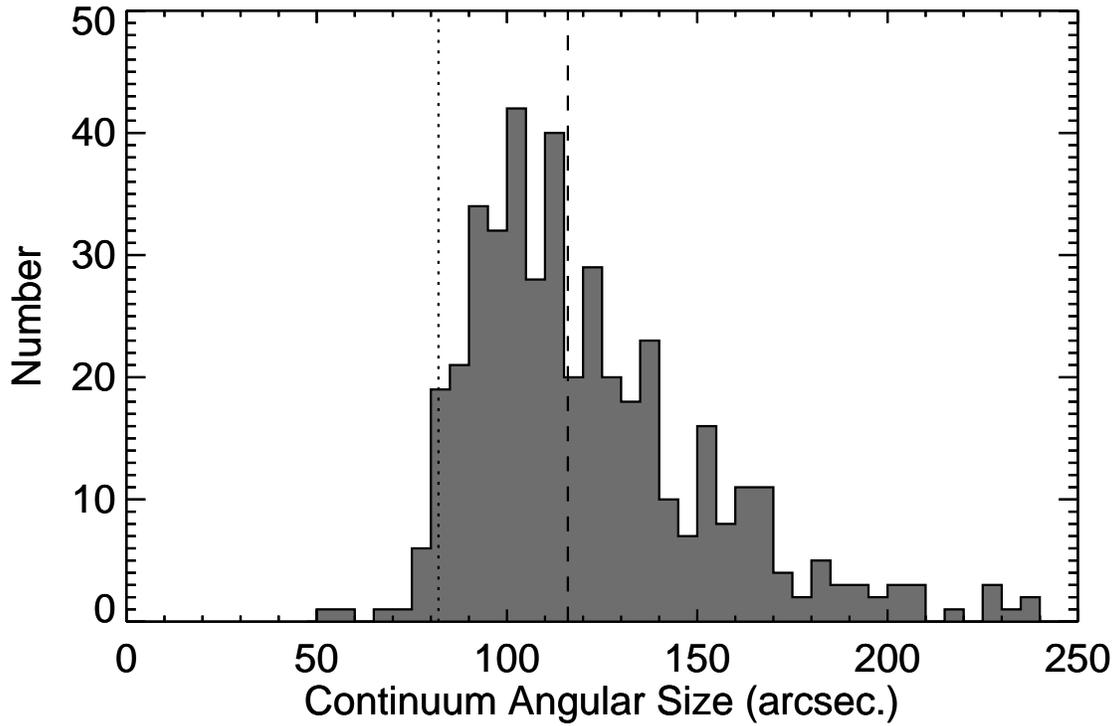}

\caption{Distribution of continuum angular size for HRDS nebulae.
Histogram shows the geometric mean of the RA and Dec angular sizes 
derived from Gaussian fits to continuum scans (see Figure~\ref{fig:example_continuum})
The vertical dotted line flags the $82\,\arcsec$ HPBW size of the
GBT beam at X-band.  The vertical dashed line flags the $116\,\arcsec$ size  
that would be measured by the GBT for a source that just fills its beam. 
}

\label{fig:continuum_size}
\end{figure}

We have constructed a web site to give others access to the HRDS
data\footnote{http://go.nrao.edu/hrds}.  On this site one can
download all the data from Tables~\ref{tab:sources}, \ref{tab:line},
and \ref{tab:continuum}, in addition to the spectra shown in
Figure~\ref{fig:example_lines}, the continuum scans shown in
Figure~\ref{fig:example_continuum}, and the three-color MIR images
shown in Figure~\ref{fig:threecolor}.  We will continue to extend this
site as more data are taken on the HRDS sources.


\begin{acknowledgments}
\nraoblurb\
We would like to thank the anonymous referee whose careful reading has
increased the clarity of this manuscript.  The HRDS was partially
supported by NSF award AST 0707853 to TMB.  LDA was partially
supported by SNF and by the NSF through GSSP awards 08-0030 and 09-005
from the NRAO.  LDA also acknowledges support from the ANR Agence
Nationale award number ANR-08-BLAN-0241.  This research has made use
of NASA's Astrophysics Data System Bibliographic Services and the
SIMBAD database operated at CDS, Strasbourg, France.

{\it Facility: Green Bank Telescope}
\end{acknowledgments}

\clearpage
\bibliographystyle{apj}

\clearpage
\begin{deluxetable}{lccccrrcc}
\tabletypesize{\scriptsize}
\tablecaption{GBT HRDS Source Catalog}
\tablewidth{0pt}
\tablehead{
\colhead{Name} &
\colhead{\gl} &
\colhead{\gb} &
\colhead{RA(J2000)} &
\colhead{Dec.(J2000)} &
\colhead{CH$_3$OH\tablenotemark{a}} &
\colhead{CS} &
\colhead{Morphology\tablenotemark{b}} &
\colhead{Note\tablenotemark{c}}
\\
\colhead{} &
\colhead{deg.} &
\colhead{deg.} &
\colhead{hh:mm:ss.s} &
\colhead{dd:mm:ss} &
\colhead{\kms} &
\colhead{\kms} &
\colhead{} &
\colhead{}
}
\startdata
G000.382$+$0.017  & $ 0.382 $ & $ +0.017 $ & 17:46:27.6 & $-$28:36:04 & \nodata & $ 84.4 $ & C & \\ 
G000.729$-$0.103  & $ 0.729 $ & $ -0.103 $ & 17:47:45.0 & $-$28:22:00 & \nodata & \nodata & I & \\ 
G000.838$+$0.189  & $ 0.838 $ & $ +0.189 $ & 17:46:52.1 & $-$28:07:20 & $ 3.5 $ & $ \phn 5.4 $ & PB & \\ 
G001.324$+$0.104  & $ 1.324 $ & $ +0.104 $ & 17:48:20.2 & $-$27:45:02 & \nodata & \nodata & PB & \\ 
G001.330$+$0.088  & $ 1.330 $ & $ +0.088 $ & 17:48:24.9 & $-$27:45:12 & \nodata & \nodata & B & \\ 
G001.330$+$0.150  & $ 1.330 $ & $ +0.150 $ & 17:48:10.2 & $-$27:43:18 & \nodata & \nodata & IB (CN25) & \\ 
G001.488$-$0.199  & $ 1.488 $ & $ -0.199 $ & 17:49:53.6 & $-$27:45:54 & \nodata & \nodata & C & \\ 
G002.009$-$0.680  & $ 2.009 $ & $ -0.680 $ & 17:52:57.8 & $-$27:33:49 & \nodata & \nodata & B (CN28) & \\ 
G002.404$+$0.068  & $ 2.404 $ & $ +0.068 $ & 17:50:58.5 & $-$26:50:32 & \nodata & \nodata & C & \\ 
G002.418$-$0.611  & $ 2.418 $ & $ -0.611 $ & 17:53:37.8 & $-$27:10:34 & \nodata & \nodata & I & \\

\enddata
\tablenotetext{a}{Methanol maser velocities are from: 
\citet{
caswell95,
ellingsen96,
gaylard93,
macleod92,
menten91,
schutte93,
slysh99,
szymczak00,
szymczak02,
vanderwalt95}
and \citet{walsh97,walsh98}.
}
\tablenotetext{b}{Classification of source structure as seen in {\it Spitzer} GLIMPSE 8\micron\ emission:\\  
B -- Bubble: 8\,\micron\ emission surrounding 24\,\micron\ and radio continuum emission\\
BB -- Bipolar Bubble:  two bubbles connected by a region of strong IR and radio continuum emission\\
PB -- Partial Bubble:  similar to ``B'' but not complete\\
IB -- Irregular Bubble:  similar to ``B'' but with less well-defined structure\\ 
C -- Compact:  resolved 8\,\micron\ emission with no hole in the center\\
PS -- Point Source:  unresolved 8\,\micron\ emission\\
I -- Irregular:  complex morphology not easily classified\\
ND -- Not Detected: source has no 8\,\micron\ emission\\
We list in parenthesis the bubble identification name for sources in the
\citet{churchwell06, churchwell07} IR bubble catalogs (see \S
\ref{sec:correlation}).}
\tablenotetext{c}{Notes on individual sources (see \S\,\ref{sec:indiv_sources}).    
PN? -- Possible Planetary Nebula (see also \S\,\ref{sec:PNe}).}
\tablecomments{Table~\ref{tab:sources} is available in its entirety in the electronic edition of the {\it Astrophysical Journal Supplement Series}. 
A portion is shown here for guidance regarding its form and content.}
\label{tab:sources}
\end{deluxetable}
\clearpage
\begin{deluxetable}{lccrcccrcc}
\tabletypesize{\scriptsize}
\tablecaption{GBT HRDS Hydrogen Recombination Line Parameters}
\tablewidth{0pt}
\tablehead{
\colhead{Name} &
\colhead{\gl} &
\colhead{\gb} &
\colhead{$T_L$} &
\colhead{$\sigma T_L$} &
\colhead{$\Delta V$} &
\colhead{$\sigma \Delta V$} &
\colhead{$V_{LSR}$} &
\colhead{$\sigma V_{LSR}$} &
\colhead{r.m.s.}
\\
\colhead{} &
\colhead{$\arcdeg$} &
\colhead{$\arcdeg$} &
\colhead{mK} &
\colhead{mK} &
\colhead{\kms} &
\colhead{\kms} &
\colhead{\kms} &
\colhead{\kms} &
\colhead{mK}
}
\startdata

G000.382$+$0.017a & 0.382 & $+$0.017 & $137.6$  &  $0.5$  &  $23.0$  &  $0.5$  &  $\phn 25.7$  &  $0.5$ & 2.4 \\
G000.382$+$0.017b & 0.382 & $+$0.017 & $104.5$  &  $2.6$  &  $32.4$  &  $2.6$  &  $\phn 41.4$  &  $2.6$ & 2.4 \\
G000.729$-$0.103a & 0.729 & $-$0.103 & $136.2$  &  $0.2$  &  $22.6$  &  $0.2$  &  $105.3$  &  $0.2$ & 2.9 \\
G000.729$-$0.103b & 0.729 & $-$0.103 & $114.4$  &  $0.4$  &  $33.1$  &  $0.4$  &  $\phn 83.2$  &  $0.4$ & 2.9\\
G000.838$+$0.189 & 0.838 & $+$0.189 & $125.2$  &  $0.1$  &  $25.4$  &  $0.1$  &  $\phn \phn 5.6$  &  $0.1$ & 2.6 \\
G001.324$+$0.104 & 1.324 & $+$0.104 & $104.7$  &  $0.1$  &  $27.0$  &  $0.1$  &  $\phn 12.7$  &  $0.1$ & 2.4\\
G001.330$+$0.088 & 1.330 & $+$0.088 & $100.2$  &  $0.1$  &  $27.4$  &  $0.1$  &  $\phn 13.2$  &  $0.1$ & 2.7\\
G001.330$+$0.150 & 1.330 & $+$0.150 & $\phn 30.6$  &  $0.1$  &  $21.7$  &  $0.1$  &  $\phn 20.1$  &  $0.1$ & 2.4 \\
G001.488$-$0.199 & 1.488 & $-$0.199 & $\phn 15.5$  &  $0.3$  &  $25.8$  &  $0.3$  &  $\phn \phn 2.2$  &  $0.3$ & 2.4\\
G002.009$-$0.680 & 2.009 & $-$0.680 & $\phn 17.3$  &  $0.3$  &  $30.2$  &  $0.3$  &  $\phn 18.2$  &  $0.3$ & 2.1\\
G002.404$+$0.068 & 2.404 & $+$0.068 & $\phn 55.8$  &  $0.1$  &  $23.1$  &  $0.1$  &  $\phn \phn 7.1$  &  $0.1$ & 2.4\\
G002.418$-$0.611 &    2.418 & $-$0.611 & $\phn \phn 8.8$  &  $0.3$  &  $18.6$  &  $0.3$  &  $\phn \phn 7.8$  &  $0.3$ & 2.0\\

\enddata
\label{tab:line}
\tablecomments{Table~\ref{tab:line} is available in its entirety in the electronic edition of the {\it Astrophysical Journal Supplement Series}. 
A portion is shown here for guidance regarding its form and content.}
\end{deluxetable}
\clearpage

\begin{deluxetable}{lccccccccccccc}
\tabletypesize{\scriptsize}
\tablecaption{GBT HRDS Radio Continuum Parameters}
\tablewidth{0pt}
\tablehead{
\colhead{Name} &
\colhead{\gl} &
\colhead{\gb} &
\colhead{$T_\alpha$} &
\colhead{$\sigma T_\alpha$} &
\colhead{$T_\delta$} &
\colhead{$\sigma T_\delta$} &
\colhead{$\Theta_\alpha$} &
\colhead{$\sigma\Theta_\alpha$} &
\colhead{$\Theta_\delta$} &
\colhead{$\sigma\Theta_\delta$} &
\colhead{$S$} &
\colhead{$\sigma S$} &
\colhead{Note\tablenotemark{a}}\\
\colhead{} &
\colhead{$\arcdeg$} &
\colhead{$\arcdeg$} &
\colhead{mK} &
\colhead{mK} &
\colhead{mK} &
\colhead{mK} &
\colhead{$\arcsec$} &
\colhead{$\arcsec$} &
\colhead{$\arcsec$} &
\colhead{$\arcsec$} &
\colhead{mJy} &
\colhead{mJy} &
\colhead{}
}
\startdata

G000.382$+$0.017 & $    0.382$  &  $   +0.017$  &  $2165$  &  $\phn 14$  &  $2239$  &  $\phn 9$  &  $105$  &  $\phn 2$  &  $\phn 88$  &  $\phn 1$  &  $1513$  &  $\phn 53$ & P\\
G000.729$-$0.103 & $    0.729$  &  $   -0.103$  &  $2335$  &  $\phn 14$  &  $2354$  &  $\phn 7$  &  $115$  &  $\phn 1$  &  $106$  &  $\phn 1$  &  $2131$  &  $\phn 55$ & \\
G000.838$+$0.189 & $    0.838$  &  $   +0.189$  &  $1025$  &  $\phn 10$  &  $1020$  &  $\phn 8$  &  $115$  &  $\phn 2$  &  $135$  &  $\phn 2$  &  $1176$  &  $\phn 73$ & \\
G001.324$+$0.104 & $    1.324$  &  $   +0.104$  &  $\phn 846$  &  $\phn \phn 5$  &  $\phn 671$  &  $\phn 8$  &  $156$  &  $\phn 2$  &  $117$  &  $\phn 2$  &  $1035$  &  $\phn 57$ & \\
G001.330$+$0.088 & $    1.330$  &  $   +0.088$  &  $\phn 869$  &  $\phn \phn 5$  &  $\phn 704$  &  $\phn 9$  &  $161$  &  $\phn 2$  &  $105$  &  $\phn 2$  &  $\phn 991$  &  $\phn 51$ & \\
G001.330$+$0.150 & $    1.330$  &  $   +0.150$  &  $\phn 137$  &  $\phn \phn 4$  &  $\phn 145$  &  $\phn 4$  &  $128$  &  $\phn 5$  &  $\phn 99$  &  $\phn 3$  &  $\phn 134$  &  $\phn 21$ & P \\
G001.488$-$0.199 & $    1.488$  &  $   -0.199$  &  $\phn 168$  &  $\phn \phn 5$  &  $\phn 171$  &  $\phn 4$  &  $131$  &  $\phn 5$  &  $135$  &  $\phn 6$  &  $\phn 223$  &  $\phn 37$ & P \\
G002.009$-$0.680 & $    2.009$  &  $   -0.680$  &  $\phn 238$  &  $\phn \phn 4$  &  $\phn 226$  &  $\phn 5$  &  $105$  &  $\phn 3$  &  $\phn 86$  &  $\phn 3$  &  $\phn 157$  &  $\phn 19$ & P \\
G002.404$+$0.068 & $    2.404$  &  $   +0.068$  &  $\phn 374$  &  $\phn \phn 8$  &  $\phn 390$  &  $19$  &  $104$  &  $\phn 3$  &  $\phn 99$  &  $\phn 7$  &  $\phn 292$  &  $\phn 67$ & P \\
G002.418$-$0.611 & $    2.418$  &  $   -0.611$  &  $\phn \phn 80$  &  $\phn \phn 3$  &  $\phn \phn 78$  &  $\phn 3$  &  $104$  &  $\phn 5$  &  $101$  &  $\phn 5$  &  $\phn \phn 61$  &  $\phn 12$ & P \\

\enddata
\tablenotetext{a}{Comments concerning continuum emission morphology and data 
quality.\\
C -- Complex:  Source has complex continuum structure; this single Gaussian 
component model only crudely represents the true source characteristics 
(see \S\,\ref{sec:radio_continuum}).\\ 
P -- Peaked:  Continuum peak lies within $10\arcsec$ of the nominal target position. 
These are the highest quality continuum data.}
\label{tab:continuum}
\tablecomments{Table~\ref{tab:continuum} is available in its entirety in the electronic edition of the {\it Astrophysical Journal Supplement Series}. 
A portion is shown here for guidance regarding its form and content.}
\end{deluxetable}
\clearpage

\begin{deluxetable}{lcll}
\tabletypesize{\scriptsize}
\tablecaption{HRDS Targets not Detected in RRL Emission}
\tablewidth{0pt}
\tablehead{
\colhead{Name} &
\colhead{Morphology\tablenotemark{a}} &
\colhead{Comments} &
\colhead{Object Type}
}
\startdata

G003.330$-$0.870	&PS	& Compact IR and radio		 			& Possible PN\\
G005.756$-$0.932	&B	& IR bubble with coincident radio 			& Unknown\\
G008.362$-$0.623	&PS	& Compact IR, coincident faint radio 			& Likely PN\\
G009.741$+$0.842	&PS	& Compact IR, coincident faint radio 			& Likely PN\\
G011.815$+$0.842	&I	& Extended bright IR, coincident faint radio		& Likely H\,{\tiny II}\\
G012.697$+$0.683	&I	& Extended IR, coincident faint radio 			& Likely H\,{\tiny II}\\
G013.751$-$0.360	&PS	& Compact IR and radio 					& Possible PN\\
G017.223$+$0.397	&PS	& Compact IR and radio					& Evolved star or PN\\
G018.446$-$0.900	&PS	& Compact IR, faint radio				& Possible PN\\
G019.236$+$0.495	&C	& Compact IR, relatively bright radio			& Possible PN\\
G038.978$-$0.268	&I	& Extended IR, Faint radio emission 			& Likely H\,{\tiny II}\\
G041.229$+$0.170	&B(N78) & IR bubble, faint radio emission 			& Likely H\,{\tiny II}\\
G045.797$-$0.347	&C	& Poor match between compact IR and diffuse radio	& Chance alignment\\
G050.312$-$0.418	&PS	& Poor match between compact IR and compact radio	& Likely AGN\\
G050.970$+$0.890	&ND	& Extended MIPSGAL, no GLIMPSE, compact radio 		& Possible AGN\\
G052.149$-$0.375	&C	& Bright compact IR, compact radio 			& Possible AGN\\
G054.096$+$0.266	&I	& Pulsar Wind Nebula associated with SNR 54.1+0.3       & PWN\\
G055.161$-$0.298	&C	& Extended GLIMPSE, faint radio 			& Likely H\,{\tiny II}\\
G064.827$+$0.494	&PS	& Poor match between compact IR and compact radio	& Likely AGN\\
G344.376$+$0.280	&PS	& Compact IR and radio	 				& Possible PN\\
G357.386$+$0.011	&ND	& Extended MIPSGAL, no GLIMPSE, faint radio 		& Evolved star or PN\\
G358.081$-$0.619	&PS	& Compact IR, faint radio 				& Possible PN\\

\enddata
\tablenotetext{a}{Source structure classification as in 
Table~\ref{tab:sources} }
\label{tab:nondetections}
\end{deluxetable}
\clearpage

\clearpage

\clearpage

\clearpage

\clearpage

\clearpage

\clearpage

\clearpage

\clearpage
\clearpage

\clearpage

\clearpage

\end{document}